\documentclass[final,notitlepage,11pt,letterpaper]{article}
\usepackage{makeidx}
\usepackage{amssymb}
\usepackage{amsmath}
\usepackage{graphicx}
\usepackage{epsfig}
\usepackage{geometry}
\usepackage{setspace}
\usepackage{endnotes}
\usepackage{rotating}
\usepackage{booktabs}
\usepackage{caption}
\usepackage{float}
\usepackage{changepage}

\graphicspath{{figs/}}

\begin{document}

\begin{onehalfspace}

\title{Fiscal Dominance and Global Determinacy}
\author{Alessandro Piergallini\thanks{%
Professor of Economics, Department of Economics and Finance, Tor Vergata
University of Rome, Via Columbia 2, 00133 Rome, Italy. ORCID:
0000-0001-6458-0364. E-mail: alessandro.piergallini@uniroma2.it. Homepage:
https://sites.google.com/view/alessandropiergallini. I would like to thank
Paolo Canofari, Giovanni Cirigliano, Alessia Franzini and Giorgio Rodano
for useful comments and discussions. Financial support
from Ministero dell'Universit\`{a} e della Ricerca (MUR), 2023--2026 PRIN
Project on \textquotedblleft Public and Corporate Debt, Monetary Policy, and
Macroeconomic Stability\textquotedblright\ (Code 2022TJWFWJ) is gratefully
acknowledged. The usual disclaimers apply.}\vspace{0.0in} \\
Tor Vergata University of Rome}
\date{October 23, 2026}
\maketitle

\vspace{-0.20in}

\begin{abstract}
\begin{adjustwidth}{-0.4in}{-0.4in}

\noindent The existing literature on monetary-fiscal policy interactions emphasizes that
an active monetary/passive fiscal policy regime causes global indeterminacy
of equilibrium, supporting liquidity traps and self-fulfilling hyperinflations as possible
equilibrium outcomes. In a nonlinear New Keynesian model with costly price adjustment, I show that
an analogous critique---with different global macroeconomic dynamics---extends
to a passive monetary/active fiscal policy regime whenever the fiscal response
to government liabilities is below the steady-state real interest rate but
positive. In this case, although the equilibrium is locally unique according
to the Blanchard-Kahn conditions, there exist multiple inflation paths
originating around the saddle path that converge asymptotically to the
central bank's target rate while satisfying the private sector's
transversality condition. I demonstrate that global determinacy instead
requires a \textit{pure} regime of fiscal dominance, with primary surpluses
set independently of the real value of government liabilities: Only then are all nonlinear
off-saddle trajectories ruled out as equilibria, for they violate the private
sector's transversality condition or its no-Ponzi game condition, or imply an
infeasible allocation. The transversality condition coincides
with the government debt valuation equation, in which primary surpluses are
discounted at the endogenous real interest rate, and pins down a unique initial
jump in inflation onto the saddle path, leading to the unique steady state.
The results imply that, when globally determinate, inflation is a fiscal
phenomenon, shaped by the central bank's interest-rate policy.
\bigskip\

 \smallskip\

 \smallskip\

\end{adjustwidth}
\end{abstract}

\thispagestyle{empty}

\end{onehalfspace}

\newpage

\pagenumbering{arabic}

\begin{doublespace}

\section{Introduction}

One influential strand of the literature on monetary-fiscal policy
interactions, initiated by Leeper (1991) and further developed by Woodford
(2001, 2003, Ch. 4.4),\footnote{%
See Canzoneri, Cumby and Diba (2010), Leeper and Leith (2016), and Bassetto
and Sargent (2020) for comprehensive reviews of the literature on the
interactions between monetary and fiscal policy, and Bianchi (2012, 2013) and
Cochrane (2022a, 2022b) for empirical interpretations of the monetary-fiscal
mix in the United States over the post-World War II period.} typically
points out the stabilizing properties of two alternative policy regimes. The
first one is characterized by an \textquotedblleft active monetary
policy\textquotedblright , which overreacts to inflationary pressures
according to the Taylor principle (Taylor, 1993, 1999),\footnote{%
See also Gal\'{\i} (2015) and Walsh (2017) for textbook treatments, and the
references therein.} and a \textquotedblleft passive fiscal
policy\textquotedblright , which \textit{per se} preserves the stability of
public debt, irrespective of the inflation path. The second regime is
characterized by an \textquotedblleft active fiscal policy\textquotedblright
, which disregards debt stabilization for any inflation trajectory, and a
\textquotedblleft passive monetary policy\textquotedblright , which
accommodates the erosion of the real value of government liabilities through
inflation to rule out debt escalation.\footnote{%
These results are known in monetary theory as the \textquotedblleft
active/passive dichotomy\textquotedblright , and prove to hold in
Ramsey-type economies with either flexible or sticky prices. Differently
from Leeper (1991), Woodford (2001, 2003, Ch. 4.4) labels a
\textquotedblleft passive (active)\textquotedblright\ fiscal policy as
\textquotedblleft locally (non-)Ricardian\textquotedblright .}

A common methodological feature of these monetary-fiscal studies---largely
relying on the workhorse dynamic stochastic general equilibrium models---is the fact that they restrict
attention to local dynamics, or small fluctuations, in the neighborhood
around the steady state in which inflation is at the target rate. This
simplification, however, has severe consequences for the issue of
macroeconomic stability. In particular, when global dynamics are taken into
account, an active monetary/passive fiscal policy causes global
indeterminacy of equilibrium, for self-fulfilling decelerating inflation
paths leading to a liquidity trap can be supported as equilibrium outcomes
(Benhabib, Schmitt-Groh\'{e} and Uribe, 2001a, 2002; Cochrane, 2017), along
with explosive inflationary paths---self-fulfilling hyperinflations---that,
depending on the model, the private sector's asymptotic conditions cannot
rule out (Cochrane, 2011, 2023, Ch. 16.3).

In a nonlinear New Keynesian model with Rotemberg-type costs of price
adjustment, I show that an analogous critique---with a different type of
global macroeconomic dynamics---extends to a passive monetary/active fiscal
policy regime whenever the response of primary surpluses to government
liabilities is below the steady-state real interest rate but remains
strictly positive, even if arbitrarily small. In this case, indeed, even
though the equilibrium is locally unique according to the commonly adopted
Blanchard-Kahn conditions (Blanchard and Kahn, 1980),\footnote{%
Such conditions explicitly account for the government's flow budget
constraint, consistently with Woodford (2001, 2003, Ch. 4.4), whereas
Benhabib, Schmitt-Groh\'{e} and Uribe (2001b) drop it under Ricardian
policies, the transversality condition being then always satisfied.} I demonstrate that there
exist multiple inflation paths originating around the saddle path that
converge asymptotically to the central bank's target rate while satisfying
the private sector's transversality condition.

In other words, once global equilibria are allowed for, a passive
monetary/active fiscal policy regime in which fiscal authorities continue to
respond to the accumulation of government liabilities by increasing---even
marginally---primary surpluses is \textit{not}, \textit{de facto}, a regime
of fiscal dominance. This is because fiscal policy, although locally
non-Ricardian, continues to be globally Ricardian, since the present
discounted value of real government liabilities converges to zero for all
possible time paths of the remaining endogenous variables.\footnote{%
That any positive feedback of primary surpluses on government liabilities
suffices for the government's intertemporal budget constraint to hold is
known from the literature on fiscal sustainability (Bohn, 1998), which takes
interest rates and prices as given and reads the result as a test of
sustainability; Cochrane (2023, pp. 139 and 143) draws the implication that,
with nominal debt, any unexpected inflation is then an equilibrium. Here the same limit condition is instead the private sector's transversality
condition---an equilibrium condition implied by optimality---and what matters
are its implications for equilibrium determination under monetary-fiscal
interactions. See also Leeper and Li
(2017), who show that single-equation surplus-debt regressions are subject to
simultaneity bias, since they neglect the forward-looking valuation of
nominal government debt and the role of monetary policy in it.} Therefore,
the paper's result, taken in conjunction with the aforementioned findings by
Benhabib, Schmitt-Groh\'{e} and Uribe (2001a) and Cochrane (2011, 2017,
2023, Ch. 16.3), yields the following remarkable implication: When fiscal
policy is globally Ricardian, monetary policy always incurs the global
indeterminacy problem, \textit{irrespective} of whether its stance is active
or passive, for solutions unrelated to fundamentals can never be ruled out.

I demonstrate that global determinacy of equilibrium requires a \textit{pure}%
\ regime of fiscal dominance, whereby fiscal policy is globally (not merely
locally) non-Ricardian---with primary surpluses set independently of the
real value of the outstanding stock of government liabilities.\footnote{%
Following the seminal contribution of Sargent and Wallace (1981), the notion
of a globally non-Ricardian regime builds on the subsequent fiscal theory of
the price level, which originates with Sims (1994), Woodford (1994, 1995),
and Cochrane (1998, 2005). See also Bassetto (2002) for a game-theoretic
foundation addressing the conceptual soundness of non-Ricardian fiscal rules.%
} The central reason is that only under such a regime do all nonlinear
off-saddle trajectories compatible with the laws of motion of macroeconomic
variables cease to be admissible equilibrium outcomes: Such
out-of-fundamentals trajectories necessarily violate either the private
sector's transversality condition or its no-Ponzi game condition, or imply
an infeasible allocation. What remains is the saddle path alone. On it, the
private sector's transversality condition coincides with the government debt
valuation equation, in which primary surpluses are discounted at the
endogenous real interest rate, and this equation pins down a unique initial
inflation rate. Under pure fiscal dominance, in other words, the government debt valuation equation \textit{is}
the saddle path; as soon as primary surpluses respond to government
liabilities, however mildly, the equivalence breaks down, and the
Blanchard-Kahn conditions select one among a continuum of equilibria. The
selected saddle path still takes the form of a valuation equation, but one
discounting future primary surpluses at the real interest rate net of the
fiscal feedback coefficient---a restriction that the private sector's
optimality conditions do not impose.

The paper's results imply that, from a global perspective, only a regime of
pure fiscal dominance, by ruling out the possibility of self-fulfilling
inflationary or disinflationary outcomes unrelated to fundamentals, is
capable of providing a nominal anchor and uniquely determining the
equilibrium paths of inflation and output, with equilibrium inflation
depending on fiscal fundamentals and on the degree of accommodation of the
central bank's interest-rate policy.

This paper complements the prominent, groundbreaking work of Cochrane (2011,
2023) in two important dimensions. The first one is the result that the
\textit{overall} local classification of active/passive regimes---not only
the one involving the application of the Taylor principle under a passive
stance of fiscal policy---is not sufficient to characterize determinacy of
equilibrium when global equilibria are taken into consideration. Cochrane (2023, Secs. 5.4.1--5.4.2) points out that, when primary surpluses
respond to the real value of government debt, however mildly, the
transversality condition holds for any unexpected inflation, so that fiscal
policy, in the terminology used here, is locally non-Ricardian but globally
Ricardian; he requires, however, a stationary debt-to-GDP ratio---a
constraint \textquotedblleft beyond the transversality condition\textquotedblright\
(p. 139)---and does not spell out this case. This paper does, in a
continuous-time New Keynesian model with sticky prices, in which the initial
price level is predetermined---initial inflation, not the price level, being
free to jump---and the Blanchard-Kahn conditions deliver a locally unique
equilibrium. When only the private sector's optimality conditions and the
feasibility of the allocation are imposed, a passive monetary/active fiscal regime
in which primary surpluses respond positively, however mildly, to the real
value of government liabilities is locally determinate but globally
indeterminate, with a continuum of equilibria that the paper characterizes
analytically.

The second dimension relates to the result that \textit{only} a
\textit{pure} regime of fiscal dominance---not merely a locally active
fiscal policy---guarantees global determinacy. Analyses of the fiscal theory with sticky prices, as in Cochrane (2023,
Ch. 5), with primary surpluses independent of the equilibrium path, select a
unique equilibrium among the stable solutions of the model by means of a
linearized government debt valuation equation, in which variations in the
discount rate enter to first order around the steady-state real interest
rate. Whether off-saddle trajectories can actually be ruled out as equilibria is,
however, a global question, whose answer depends on endogenous discounting,
on how primary surpluses respond to government liabilities, and on the
productive capacity of the economy. This paper therefore addresses the question of equilibrium uniqueness from a
global perspective. It starts
from the conventional Blanchard-Kahn saddle-path apparatus and asks: Under
which locally stabilizing configurations of the monetary-fiscal policy mix
are all nonlinear off-saddle trajectories inadmissible, that is,
inconsistent with either the private sector's asymptotic conditions or the
feasibility of the allocation they imply? Only under such configurations
does local determinacy \textit{extend} to global determinacy. The key reason is that pure fiscal dominance is the one configuration under
which the transversality condition discriminates among candidate paths.
With primary surpluses set independently of the real value of government
liabilities, it becomes a genuine restriction---the government debt valuation
equation---which holds on the saddle path and fails off it. Under a positive fiscal feedback, however small, primary surpluses rise with
government liabilities and ensure that the present discounted value of the
latter vanishes along every path: The transversality condition is then always
satisfied, and the valuation equation restricts nothing. Global determinacy,
in this sense, rests on the same instrument invoked by Cochrane (2011, 2023),
but only \textit{once} the nonlinear dynamics of the model have restricted the
admissible equilibrium candidates to the saddle path alone.
When globally determinate, inflation is thus a fiscal phenomenon, shaped by
the central bank's interest-rate policy.

The global analysis also sheds new light on the explosive inflationary paths
that, according to Cochrane (2011), nothing in the private sector's
asymptotic conditions can rule out under an active stance of monetary
policy. Indeed, the transversality condition alone does not exclude them:
Along such paths the real interest rate rises without bound, and real
government liabilities are discounted away. In the nonlinear New Keynesian economy studied here, however, these paths
cannot occur, for an ever-rising real interest rate would require
consumption, and thus output, to grow without bound over time, which is
incompatible with the economy's productive capacity.
The oscillating explosive solutions of the linearized model end, globally,
either in the liquidity trap or in a deflationary collapse that is not an
equilibrium either. The global pathology of an active monetary/passive
fiscal regime is therefore the liquidity trap, and it survives intact,
including equilibrium paths that originate arbitrarily close to the target.
Differently from Benhabib, Schmitt-Groh\'{e} and Uribe (2001a), where such
paths are one of two possible outcomes---the other being convergence to a
limit cycle---and are established near a bifurcation point, the absence of
closed orbits makes them here the only possible outcome, for all admissible
parameter values.%
\footnote{%
Kaplan (2026) studies equilibrium uniqueness in New Keynesian models by
combining boundedness of the real block with the transversality condition on
government debt, under bounds on inflation. In the present framework, the
exclusion of explosive inflationary paths is derived rather than assumed:
Under passive monetary policy, from the private sector's no-Ponzi game
condition; under active monetary policy, from the feasibility of the implied
allocation.}

It should be stressed that the argument by which pure fiscal dominance
delivers global determinacy does not rely on a policy \textquotedblleft
threat to blow up the economy\textquotedblright\ (Cochrane, 2011, p. 567):
That is, a fiscally unsustainable trigger policy that is activated should the economy threaten
to abandon the target steady state and slide into an undesirable
equilibrium, such as a liquidity trap (Benhabib, Schmitt-Groh\'{e} and
Uribe, 2002). According to Cochrane (2011), such a policy threat would
introduce a questionable game between the atomistic private sector and the
government. The paper's point, rather, hinges on the fundamental reason why,
in an environment with forward-looking optimizing agents, jumping variables
must stay on the saddle path under a \textit{given}, time-invariant policy
configuration of the monetary-fiscal mix, with no announced contingent
switch: \textit{All other} initial values would lead the economy onto paths
that either necessarily violate the asymptotic conditions characterizing
private dynamic optimization or cannot be sustained by the economy's
resources.

To make the argument as direct and transparent as possible, Section 2 sets
up a cashless New Keynesian model in continuous time, which deliberately
excludes stochastic shocks, and characterizes the perfect-foresight
equilibrium. Section 3 analyzes the issue of global dynamics and proves that
global determinacy does not hold whenever fiscal policy remains globally
Ricardian---as is the case not only under the canonical active
monetary/passive fiscal regime, but also under a passive monetary/active
fiscal regime whenever the response of primary surpluses to government
liabilities remains strictly positive: Only a regime of pure fiscal
dominance, in which fiscal policy is globally non-Ricardian, is capable of
guaranteeing global uniqueness of equilibrium. Section 4 discusses the implications of the results and provides concluding
remarks. Formal proofs are collected in the Appendices.

\section{The Model}

To convey the paper's main arguments in a direct and intuitive way by means
of phase diagrams, I employ a cashless, nonlinear New Keynesian model set in
continuous time, capable of generating an autonomous system of differential
equations in output, inflation, and real government debt. Nominal rigidities
are modeled through quadratic costs of price adjustment \`{a} la Rotemberg
(1982), in the specification of Benhabib, Schmitt-Groh\'{e} and Uribe
(2001a), in which the costs of changing prices enter the utility function of
household-firms. Following Woodford (2003) and Cochrane (2011, 2023), money
plays no role other than that of unit of account: The economy is cashless,
and government liabilities consist of nominal bonds only.\footnote{%
Under a money-in-the-utility-function specification \`{a} la Sidrauski
(1967) and Brock (1974, 1975) with preferences separable in consumption and
real balances, the money market equilibrium condition would only determine
the quantity of money residually, the interest rate being the policy
instrument. Real money balances would enter total government liabilities,
and seigniorage the primary surplus---two terms that vanish in the cashless
limit (Woodford, 2003). Under nonseparable preferences, real balance effects
on the marginal utility of consumption have been found to be quantitatively
small in calibrated and estimated models (e.g., McCallum, 2001; Woodford,
2003; Ireland, 2004; Andr\'{e}s, L\'{o}pez-Salido and Vall\'{e}s, 2006),
and, as Woodford (2008) argues, the analysis of interest-rate policy can abstract from monetary aggregates.}

The choice of Rotemberg rather than Calvo (1983) pricing is dictated by the
global nature of the analysis. The two specifications are equivalent to a
first-order approximation (Roberts, 1995), and Benhabib, Schmitt-Groh\'{e}
and Uribe (2001b, Appendix) show that local determinacy results carry over
from one to the other. Away from the steady state, however, Calvo pricing
introduces price dispersion as an additional state variable, so that the
nonlinear dynamics cannot be reduced to a Phillips curve in inflation alone
(Ascari and Rossi, 2012; Ascari and Sbordone, 2014). Under Rotemberg pricing,
all firms set the same price in equilibrium, the nonlinear Phillips curve is
exact, and the real block of the model remains two-dimensional, which is what
makes a global phase-plane analysis possible.\footnote{%
Whether the global results extend to Calvo pricing, with price dispersion as
an additional state variable, remains an open question.}

Two key consequences for the purposes of this paper arise. First, since
prices are sticky and time is continuous, the initial price level is predetermined, and so is the
initial real stock of government debt. The government debt valuation equation \`{a} la Cochrane (2011, 2023) thus
becomes an equation in initial inflation---not in the initial price
level---while still delivering, as in the analysis of sticky prices in
Cochrane (2023, Ch. 5), a fiscal theory of inflation. Second, and most importantly, since the real interest rate responds to
inflation through the stance of monetary policy and moves along the
adjustment path, the discount rate in the government debt valuation equation
is endogenous, as in Cochrane (2023, Ch. 5). Differently from his linearized
analysis, however, the valuation equation must here be retained in its exact
nonlinear form: Along off-saddle trajectories, variations in the discount
rate are not small, and they are precisely what, globally, determines whether
such trajectories can be ruled out.

\subsection{The Private Sector}

The economy consists of a continuum of identical, infinitely lived
household-firms, each producing a differentiated good with labor as the
only input. Household-firm $j$ derives utility from consumption of a
composite good, $c$, disutility from working, $h$, and disutility from
changing the price of its own good, $P_{j}$, at a rate different from the
inflation target. Its lifetime utility is given by%
\begin{equation}
U(0) =\int_{0}^{\infty }e^{-rt}\left[ u(c)-z(h)-\frac{\gamma }{2}%
\left( \frac{\dot{P}_{j}}{P_{j}}-\pi ^{\ast }\right) ^{2}\right] dt,
\label{uf}
\end{equation}%
where $r>0$ denotes the rate of time preference, $\gamma >0$ measures the
degree of price stickiness, and $\pi ^{\ast }$ is the central bank's
inflation target. The instant utility function $u(\cdot) $ is
twice continuously differentiable, strictly increasing, and strictly
concave, satisfies the Inada conditions, and exhibits a coefficient of
relative risk aversion $\sigma (c) \equiv -cu^{\prime \prime
}(c) /u^{\prime }(c) \geq 1$.\footnote{%
The restriction $\sigma \geq 1$ is sufficient, not necessary, for the
monotonicity of the Phillips curve established below. In the steady state
the relevant monotonicity holds for any $\sigma >0$.} The disutility of
labor $z(\cdot) $ is defined on $\left[ 0,\bar{h}\right) $,
where $\bar{h}$ denotes the endowment of time, and is strictly increasing
and strictly convex, with $z^{\prime }(0) <\infty $ and $%
\lim_{h\rightarrow \bar{h}}z^{\prime }(h) =\infty $. This
standard Inada-type property---leisure is essential---ensures that hours
can never exhaust the time endowment. Output is produced according to the
technology $y_{j}=f(h_{j}) $, where $f(\cdot) $ is
strictly increasing and concave, with $f(0) =0$ and $%
\lim_{h\rightarrow 0}f(h) /f^{\prime }(h) =0$. Let $%
\bar{y}\equiv f(\bar{h}) $ denote the productive capacity of the
economy.

The demand for good $j$ is given by $y_{j}=(P_{j}/P) ^{\eta }y$,
where $P$ denotes the price index, $y$ aggregate demand, and $\eta <-1$ the
price elasticity of demand. Letting $B$ denote nominal government bonds,
which pay the nominal interest rate $R$, $b\equiv B/P$ the real stock of
bonds, $\pi \equiv \dot{P}/P$ the inflation rate, and $\tau $ real lump-sum
taxes net of transfers, the instant budget constraint of household-firm $j$ formulated in real
terms is given by%
\begin{equation}
\dot{b}=(R-\pi) b+\frac{P_{j}}{P}f(h_{j}) -c-\tau .
\label{hbc}
\end{equation}
The household-firm chooses time paths for $c$, $h_{j}$, $P_{j}$, and $b$ so
as to maximize (\ref{uf}) subject to (\ref{hbc}), the demand constraint $%
f(h_{j}) =(P_{j}/P) ^{\eta }y$, the initial
conditions $B(0) =B_{0}$ and $P_{j}(0) =P_{0}$, and
the no-Ponzi game condition%
{\setlength{\abovedisplayskip}{2pt}
\begin{equation}
\underset{t\rightarrow \infty }{\lim }e^{-\int_{0}^{t}\left[ R(v)-\pi (v)%
\right] dv}b(t) \geq 0,  \label{npg}
\end{equation}}%
taking the paths of lump-sum taxes, nominal interest rates, inflation rates,
and aggregate demand as parametrically given. Letting $\lambda $ denote the
co-state variable associated with $b$, optimality yields%
\begin{equation}
u^{\prime }(c)=\lambda ,  \label{uc}
\end{equation}%
\begin{equation}
R-\pi =r-\frac{\dot{\lambda}}{\lambda },  \label{cs10}
\end{equation}%
an optimal pricing condition, derived in Appendix A, and the
transversality condition%
\begin{equation}
\underset{t\rightarrow \infty }{\lim }e^{-\int_{0}^{t}\left[ R(v)-\pi (v)%
\right] dv}b(t) =0,  \label{tvc}
\end{equation}%
which asserts that, because utility is strictly increasing in consumption,
at the optimum the household-firm will satisfy the borrowing limit (\ref{npg}%
) with equality.\footnote{%
Since, from (\ref{cs10}), $\lambda (t) =\lambda (0)
e^{rt-\int_{0}^{t}\left[ R(v)-\pi (v)\right] dv}$, condition (\ref{tvc}) can
be equivalently written as $\lim_{t\rightarrow \infty }e^{-rt}\lambda (t) b(t) =0$. No transversality condition on the price
choice, nor any a priori bound on inflation, is imposed.}

Since prices are sticky, the price level cannot jump: $P(0)
=P_{0}$. As a consequence, the initial real stock of government debt%
\begin{equation}
b(0) =\frac{B_{0}}{P_{0}}  \label{ic}
\end{equation}%
is predetermined, whereas consumption and inflation are free to jump at time
zero.

\subsection{The Public Sector}

Monetary policy is described in terms of an interest-rate feedback rule,
whereby the nominal interest rate controlled by monetary authorities is set
as an increasing function of the inflation rate. Then,%
\begin{equation}
R=R(\pi) ,  \label{mpr}
\end{equation}%
where function $R(\cdot) $ is continuously differentiable,
strictly increasing, strictly convex, and strictly positive, with $%
\lim_{\pi \rightarrow -\infty }R(\pi) =0$: The nominal interest
rate approaches its zero lower bound asymptotically as deflation deepens,
without ever reaching it,  as in Benhabib,
Schmitt-Groh\'{e} and Uribe (2001a).\footnote{%
In a cashless economy the zero lower bound does not arise from the presence
of money in the household's portfolio but is a property of the policy rule,
as in the cashless limit studied by Woodford (2003) and in the literature on
liquidity traps in cashless New Keynesian models (e.g., Eggertsson and
Woodford, 2003; Werning, 2011; Cochrane, 2017).} The central bank is assumed
to target an inflation rate $\pi ^{\ast }>-r$ that satisfies $R(\pi
^{\ast }) =r+\pi ^{\ast }$.

Following the terminology of Leeper (1991),y, monetary policy is referred to as
\textquotedblleft active\textquotedblright\ if $R^{\prime }(\pi) >1$. That is, under an active stance of interest rate policy,
whenever monetary authorities observe symptoms of inflationary pressure,
they will tighten policy sufficiently to ensure an increase in the real rate
of interest, in accordance with the Taylor principle (Taylor, 1993, 1999).
Conversely, monetary policy is \textquotedblleft passive\textquotedblright\
if $R^{\prime }(\pi) <1$, that is,\ whenever the central bank
underreacts to inflationary pressures, inducing a decrease in the real
interest rate.

Two configurations of the policy rule will be considered. In the first, the
rule satisfies the Taylor principle around the target, $R^{\prime }(\pi ^{\ast }) >1$. Since $R(\cdot) $ is convex and
approaches zero as $\pi \rightarrow -\infty $, the Fisher relation $R(\pi) =r+\pi $ is then satisfied not only at the target but also at an
unintended inflation rate $\pi _{L}<\pi ^{\ast }$, with $R^{\prime }(\pi _{L}) <1$, in the vicinity of the zero lower bound.\footnote{%
The exponential rule $R(\pi) =(r+\pi ^{\ast }) \exp
\left\{ \left[ A/(r+\pi ^{\ast }) \right] (\pi -\pi ^{\ast
}) \right\} $, with $A>1$, employed by Benhabib, Schmitt-Groh\'{e} and
Uribe (2001a), satisfies all the above properties.} In the second, the rule
is globally passive: Besides approaching zero as $\pi \rightarrow -\infty $,
it is asymptotic, as $\pi \rightarrow \infty $, to a straight line with slope
$\varphi <1$. Convexity then implies that $R^{\prime }(\pi) $
increases from zero to $\varphi $, so that $0<R^{\prime }(\pi)
<\varphi <1$ for every $\pi $. The real interest rate $R(\pi)
-\pi $ is then strictly decreasing in inflation, from $+\infty $ to $-\infty
$, so that the target is the only inflation rate satisfying the Fisher
relation. That the asymptotic slope of the rule be strictly lower than unity
will prove essential: It implies that, should inflation diverge, the real
interest rate would fall without bound.

Government purchases are set equal to zero, for simplicity and without loss
of generality. Public deficits are financed by issuing nominal bonds. Then,
letting $s\equiv \tau $ denote the real primary surplus, the instant budget
constraint of the public sector can be expressed in terms of the law of
motion of the real stock of government debt, given by
\begin{equation}
\dot{b}=(R-\pi) b-s.  \label{gibc}
\end{equation}%
To close the model, one needs to specify the behavior of fiscal policy.
Following Leeper (1991) and Bohn (1995, 1998), I consider a debt-based
primary-surplus feedback rule of the form%
\begin{equation}
s=\bar{s}+\alpha b,  \label{fpr}
\end{equation}%
where $\bar{s}$ is a bounded, time-varying exogenous term and $\alpha \geq 0$
a constant.\footnote{%
Government debt consists of short-term nominal bonds, so that the comparison
is with the short-term debt case in Cochrane (2023), who also considers
long-term debt.}

Fiscal policy is referred to as \textquotedblleft passive\textquotedblright\
(following the terminology of Leeper, 1991) or \textquotedblleft locally
Ricardian\textquotedblright\ (following the terminology of Woodford, 2001,
2003, Ch. 4.4) if, when substituting the budgetary policy rule (\ref{fpr})
into the government's flow budget constraint (\ref{gibc}), public debt is
locally stable in the neighborhood around a stationary state---whereby $\dot{%
b}=0$ and $\dot{\lambda}=0$---for any small enough fluctuations in the
remaining endogenous variables. Formally, using the co-state equation (\ref%
{cs10}), this occurs if%
\begin{equation}
\left. \frac{\partial \dot{b}}{\partial b}\right\vert _{\dot{b}=0\text{, }%
\dot{\lambda}=0}=r-\alpha <0,  \label{pfp}
\end{equation}%
that is, if the degree of fiscal reactiveness to the accumulation of
government liabilities is higher than the steady-state real interest rate, $%
\alpha >r$. Conversely, fiscal policy is \textquotedblleft
active\textquotedblright\ or \textquotedblleft locally
non-Ricardian\textquotedblright\ if it does not preserve \textit{per se}
local stability of public debt, that is, if%
\begin{equation}
\left. \frac{\partial \dot{b}}{\partial b}\right\vert _{\dot{b}=0\text{, }%
\dot{\lambda}=0}=r-\alpha >0,  \label{afp}
\end{equation}%
implying that the feedback fiscal policy parameter is lower than the
steady-state real interest rate, $0\leq \alpha <r$.

\subsection{Equilibrium}

I am now in the position to characterize a perfect-foresight equilibrium.
Since the costs of price adjustment are expressed in utility terms,
combining the private and public budget constraints (\ref{hbc}) and (\ref%
{gibc}) in a symmetric equilibrium, in which $P_{j}=P$ for all $j$, yields
the product market equilibrium condition%
\begin{equation}
c=y=f(h) <\bar{y}.  \label{pm-equilibrium}
\end{equation}%
Using (\ref{pm-equilibrium}) into (\ref{uc}) and differentiating with
respect to time, the co-state equation (\ref{cs10}) gives rise to the
dynamic IS curve%
\begin{equation}
\dot{y}=\frac{y}{\sigma (y) }\left[ R(\pi) -\pi -r%
\right] ,  \label{IS}
\end{equation}%
according to which output grows whenever the real interest rate exceeds the
rate of time preference. The optimal pricing condition, in turn, gives rise
to the nonlinear New Keynesian Phillips curve (see Appendix A for the
derivation)%
\begin{equation}
\dot{\pi}=r(\pi -\pi ^{\ast }) -\Gamma (y) ,
\label{NKPC}
\end{equation}%
where%
\begin{equation}
\Gamma (y) \equiv \frac{y}{\gamma }\left[ (1+\eta)
u^{\prime }(y) -\eta \frac{z^{\prime }(f^{-1}(y)) }{f^{\prime }(f^{-1}(y)) }\right] =%
\frac{\left\vert \eta \right\vert yu^{\prime }(y) }{\gamma }%
\left[ mc(y) -\frac{1+\eta }{\eta }\right]   \label{Gamma}
\end{equation}%
is proportional to the gap between the real marginal cost and its
steady-state value, the inverse of the markup $\eta /(1+\eta) $%
, the real marginal cost being given by $mc(y) \equiv z^{\prime
}(f^{-1}(y)) /\left[ u^{\prime }(y)
f^{\prime }(f^{-1}(y)) \right] $. As shown in Appendix B, $\Gamma (\cdot) $ is strictly increasing, with $\Gamma (y) \rightarrow
+\infty $ as $y\rightarrow \bar{y}$---because hours approach the time
endowment, at which the marginal disutility of labor explodes---and $\Gamma (y) \rightarrow \Gamma _{0}<0$ as $y\rightarrow
0$, where $\Gamma _{0}\equiv \lim_{y\rightarrow 0}\Gamma (y) $
is finite under logarithmic utility and equal to $-\infty $ when $\sigma >1$. The steady-state
level of output $y^{\ast }$ is the unique solution of $\Gamma (y) =0$, and $\kappa \equiv \Gamma ^{\prime }(y^{\ast }) >0$
denotes the slope of the Phillips curve. Integrating (\ref{NKPC}) forward,
along every equilibrium characterized below, the Phillips curve takes the
familiar form%
\begin{equation*}
\pi (t) -\pi ^{\ast }=\int_{t}^{\infty }e^{-r(v-t)
}\Gamma (y(v)) dv,
\end{equation*}%
according to which current inflation reflects the present discounted value
of future marginal-cost gaps.

Using the monetary and fiscal policy rules (\ref{mpr}) and (\ref{fpr}) to
eliminate $R$ and $s$ from (\ref{gibc}) yields the law of motion of real
government debt%
\begin{equation}
\dot{b}=\left[ R(\pi) -\pi -\alpha \right] b-\bar{s}.
\label{ODE-b}
\end{equation}%
Equations (\ref{IS}), (\ref{NKPC}), and (\ref{ODE-b}) form an autonomous,
nonlinear differential-equation system in the variables $y$, $\pi $, and $b$
(when $\bar{s}$ is constant), with a recursive structure: The real
block---the dynamic IS curve (\ref{IS}) and the Phillips curve (\ref{NKPC}%
)---does not depend on real government debt, whose law of motion (\ref%
{ODE-b}) in turn depends on the real block only through the equilibrium real
interest rate $R(\pi) -\pi $.

A perfect-foresight competitive equilibrium is a set of time paths $\left\{
y,\pi ,b\right\} $, defined for all $t\geq 0$ with $0<y<\bar{y}$, satisfying (%
\ref{ic}), (\ref{IS}), (\ref{NKPC}), (\ref{ODE-b}), and the private sector's
transversality condition written as%
\begin{equation}
\underset{t\rightarrow \infty }{\lim }e^{-\int_{0}^{t}\left[ R(\pi
(v)) -\pi (v)\right] dv}b(t) =0,  \label{TVC-equilibrium}
\end{equation}%
given the initial conditions $B(0) =B_{0}$ and $P(0)
=P_{0}$.

\section{Global Dynamics}

At this point I am equipped with the elements to demonstrate the paper's
main point: The local determinacy properties of the equilibrium induced by
\textit{both} an active monetary/passive fiscal policy regime \textit{and} a
passive monetary/active fiscal policy regime vanish as soon as one adopts a
global-dynamics perspective whenever fiscal authorities maintain a \textit{%
positive}---even if arbitrarily small in the latter regime---feedback
reaction to the accumulation of government liabilities ($\alpha >0$); and
such properties can survive globally---thus preserving their general
validity---only under a \textit{purely} fiscal-dominant framework, whereby
the budgetary policy stance is globally---not merely locally---non-Ricardian
($\alpha =0$).

Two general properties of the dynamic system (\ref{IS})-(\ref{NKPC})-(\ref%
{ODE-b}) will be used throughout. The first concerns the counting of
variables. Since the initial real stock of government debt $b(0) $
is predetermined, while output and inflation are jump variables, the
Blanchard-Kahn conditions require exactly one stable eigenvalue of the
Jacobian matrix evaluated at the steady state. Because of the recursive
structure of the system, this matrix is block-triangular, and its
eigenvalues are those of the real block together with%
\begin{equation*}
\left. \frac{\partial \dot{b}}{\partial b}\right\vert _{\dot{b}=0\text{, }%
\dot{y}=0}=r-\alpha ,
\end{equation*}%
which is negative under passive and positive under active fiscal policy.
Linearizing the real block around a steady state $(\tilde{y},\tilde{\pi}) $ yields the Jacobian matrix%
\begin{equation}
J^{(\tilde{y},\tilde{\pi}) }=\left(
\begin{array}{cc}
0 & \frac{\tilde{y}}{\sigma (\tilde{y}) }\left[ R^{\prime }(\tilde{\pi}) -1\right]  \\
-\Gamma ^{\prime }(\tilde{y})  & r%
\end{array}%
\right) ,  \label{jacobian}
\end{equation}%
whose trace equals the rate of time preference $r>0$ and whose determinant,%
\begin{equation*}
\det J^{(\tilde{y},\tilde{\pi}) }=\Gamma ^{\prime }(\tilde{y})
\frac{\tilde{y}}{\sigma (\tilde{y}) }\left[ R^{\prime }(\tilde{\pi}) -1\right] ,
\end{equation*}%
takes the sign of $R^{\prime }(\tilde{\pi}) -1$. The real block thus
exhibits a saddle point wherever monetary policy is locally passive and a
source wherever it is locally active.

The second property is global. In logarithmic coordinates for the marginal
utility of consumption, the divergence of the vector field governing the
real block is constant and equal to the rate of time preference, so that, by
Bendixson's criterion, no closed orbits---and no closed chains of
trajectories connecting steady states---exist (see Appendix C). Every
trajectory of the real block either converges to a steady state or diverges.%
\footnote{%
This property distinguishes the cashless specification from the
money-in-the-utility-function specification with nonseparable preferences
employed by Benhabib, Schmitt-Groh\'{e} and Uribe (2001a), in which the
divergence varies along trajectories and periodic equilibria may arise.}

To render the argument as transparent as possible, it is first worth
scrutinizing what happens in the present framework in a regime of monetary
dominance whereby monetary policy is active and fiscal policy is passive.

\subsection{Active Monetary/Passive Fiscal Policy Regime}

As commonly advocated in the standard monetary analysis since Taylor (1993),
consider the case of a monetary policy reaction function satisfying the
Taylor principle around the inflation target, $R^{\prime }(\pi ^{\ast
}) >1$, in conjunction with a fiscal policy reaction function setting $%
\alpha >r$, thus guaranteeing stability of real public debt for any small
enough fluctuation in inflation.

Consider first the real block. Setting $\dot{y}=0$ in equation (\ref{IS})
yields the Fisher equation%
\begin{equation}
R(\pi) =r+\pi .  \label{pi-locus}
\end{equation}%
In the phase plane $(y,\pi) $, the $\dot{y}=0$-locus is thus
horizontal in correspondence of \textit{both} the target inflation rate $\pi
^{\ast }$ \textit{and} the unintended, possibly negative, inflation rate $%
\pi _{L}<\pi ^{\ast }$ in the vicinity of the zero lower bound on the
nominal interest rate. In addition, from (\ref{IS}),%
\begin{equation}
\dot{y}>(<) 0\text{ \ \ if \ \ }R(\pi) >(<) r+\pi ,  \label{y-dot}
\end{equation}%
implying $\dot{y}>(<) 0$ whenever $\pi >\pi ^{\ast }$ or $\pi
<\pi _{L}$ ($\pi _{L}<\pi <\pi ^{\ast }$): Output rises whenever the real
interest rate exceeds the rate of time preference. On the other hand,
setting $\dot{\pi}=0$ in equation (\ref{NKPC}) yields the long-run Phillips
curve%
\begin{equation}
\pi =\pi ^{\ast }+\frac{\Gamma (y) }{r},  \label{y-locus}
\end{equation}%
which is upward sloping and exhibits a vertical asymptote in correspondence
of the productive capacity $\bar{y}$: Along it, higher steady-state
inflation is associated with higher output. From (\ref{NKPC}), $\dot{\pi}%
>(<) 0$ above (below) the $\dot{\pi}=0$-locus.

The two loci intersect at the two steady states $(y^{\ast },\pi ^{\ast
}) $ and $(y_{L},\pi _{L}) $, where $y_{L}<y^{\ast }$
solves $\Gamma (y_{L}) =r(\pi _{L}-\pi ^{\ast }) <0$%
: The unintended steady state features lower inflation and lower output.%
\footnote{%
The unintended steady state exists if $\Gamma _{0}<r(\pi _{L}-\pi
^{\ast }) $. Since $\Gamma _{0}=-\left( \left\vert 1+\eta \right\vert
/\gamma \right) \lim_{y\rightarrow 0}yu^{\prime }(y) $, the
condition is automatically satisfied whenever $\sigma >1$, in which case $%
yu^{\prime }(y) \rightarrow \infty $ as $y\rightarrow 0$, and
reduces to $\left\vert 1+\eta \right\vert /\gamma >r(\pi ^{\ast }-\pi
_{L}) $ under logarithmic utility.} A steady state of the full system
requires the exogenous part of the real primary surplus to be stationary, $%
\bar{s}(t) =\bar{s}^{\ast }$; whenever convergence to a steady
state is considered, $\bar{s}(t) $ is assumed to converge to $%
\bar{s}^{\ast }$. Since both $\pi ^{\ast }$ and $\pi
_{L}$ solve $R(\pi) -\pi =r$, equation (\ref{ODE-b}) delivers
at both steady states the same stock of real government debt%
\begin{equation}
b^{\ast }=\frac{\bar{s}^{\ast }}{r-\alpha },  \label{SSb}
\end{equation}%
which, since fiscal policy is passive, is positive if the exogenous
component of the primary surplus is in deficit, $\bar{s}^{\ast }<0$.

\begin{figure}[!htb]
\centering
\includegraphics[width=0.78\textwidth]{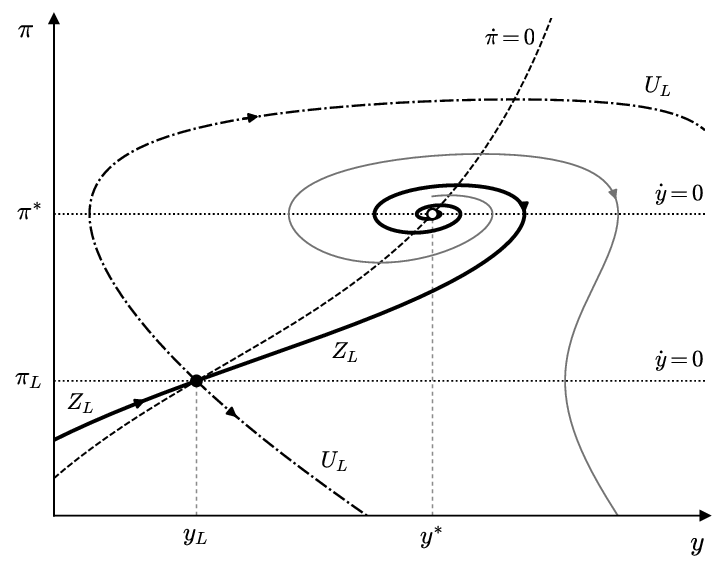}
\caption{Global Dynamics under an Active Monetary/Passive Fiscal Policy Regime}\medskip
\begin{minipage}{1.0\textwidth}
\footnotesize \textit{Notes}: The figure depicts the phase plane $(y,\pi) $ of the real
block. The dashed curve is the long-run Phillips curve $\dot{\pi}=0$; the
dotted lines are the Fisher relations $\dot{y}=0$ at $\pi ^{\ast }$ and $\pi
_{L}$. The intended steady state $(y^{\ast },\pi ^{\ast }) $
(empty circle) is a source; the liquidity trap $(y_{L},\pi _{L}) $
(full circle) is a saddle, with stable arm $Z_{L}Z_{L}$ (solid) and unstable
arm $U_{L}U_{L}$ (dash-dotted). One branch of $Z_{L}Z_{L}$ spirals out of
the intended steady state (heteroclinic connection). The grey trajectory
starts close to the target and, after widening oscillations, ends in a
deflationary collapse that is not an equilibrium. The figure is drawn from a
numerical illustration (see Appendix G).
\end{minipage}
\end{figure}

Making use of the foregoing properties of the dynamic system gives rise to
the global dynamics characterized in Figure 1. Around the intended steady
state, the real block is a source: The Jacobian matrix (\ref{jacobian})
evaluated at $(y^{\ast },\pi ^{\ast }) $ has positive trace and
positive determinant, so that both its eigenvalues---complex conjugates whenever $4\kappa
\left[ y^{\ast }/\sigma (y^{\ast }) \right] \left[ R^{\prime
}(\pi ^{\ast }) -1\right] >r^{2}$, as in the case
depicted---have positive real parts, while the eigenvalue associated with
real government debt, $r-\alpha $, is negative. Since $b$ is
a state variable and $y$ and $\pi $ are jump variables, the Blanchard-Kahn
condition is satisfied, and there exists around the steady state a unique
equilibrium trajectory converging asymptotically to it. Specifically, for a
given $b(0) $, the only trajectory converging asymptotically to $%
(y^{\ast },\pi ^{\ast },b^{\ast }) $ is given by the saddle-path
solution%
\begin{equation}
y(t) =y^{\ast },\qquad \pi (t) =\pi ^{\ast },
\label{ZZ1-AM/PF}
\end{equation}%
\begin{equation}
b(t) =b^{\ast }+e^{-(\alpha -r) t}(b(0) -b^{\ast }) -\int_{0}^{t}e^{-(\alpha -r) (t-v) }\left[ \bar{s}(v) -\bar{s}^{\ast }\right] dv.\medskip
\label{ZZ2-AM/PF}
\end{equation}%
In the absence of stochastic shocks, if one restricts attention to local
dynamics, according to (\ref{ZZ1-AM/PF}) the central bank is perfectly able
to pin down inflation at the target rate, and output at its steady-state
level, by increasing the nominal interest rate by more than one-to-one in
response to potential inflationary pressures, while according to (\ref%
{ZZ2-AM/PF}) real government debt converges to the steady state at a
velocity proportional to the gap between the primary-surplus feedback
reaction to the accumulation of government liabilities implemented by
fiscal authorities and the steady-state real interest rate. Deviations of
the exogenous primary surplus from its stationary value are absorbed by real
government debt and leave inflation and output unaffected: Debt reflects
past, not future, primary surpluses. Since prices are sticky, the initial price level is predetermined, and the
nominal indeterminacy that would affect this regime under flexible
prices---the initial price level being left unpinned by the Taylor
principle (see Benhabib, Schmitt-Groh\'{e} and Uribe, 2001b)---does not
arise here. The description of the global dynamics that follows refers to the case of
complex eigenvalues, which obtains unless the Taylor principle is only barely
satisfied or the Phillips curve is very flat; with real eigenvalues,
trajectories leave the intended steady state without oscillating, but none
of the results is affected.

Once global dynamics are explored, however, a fundamental pathology emerges,
generating serious cases of macroeconomic instability. Consistently with the
point originally made by Benhabib, Schmitt-Groh\'{e} and Uribe (2001a),
consider the unintended steady state $(y_{L},\pi _{L}) $. Since $%
R^{\prime }(\pi _{L}) <1$, the Jacobian matrix (\ref{jacobian})
evaluated there has a negative determinant: The unintended steady state is a
saddle point of the real block, with a stable arm, $Z_{L}Z_{L}$, and an
unstable arm, $U_{L}U_{L}$. Contrary to the case in which a predetermined
variable pins down the initial point on the stable arm of a saddle, both
output and inflation are here free to jump. As a consequence, \textit{every}
point of $Z_{L}Z_{L}$ is an admissible initial condition, from which output
and inflation converge to $(y_{L},\pi _{L}) $. There exists a
continuum of self-fulfilling decelerating inflation paths converging to an
unintended low-inflation steady state that exhibits the features of a
liquidity trap, in the sense that monetary policy is powerless to reverse
the dynamics and, by cutting the nominal interest rate aggressively as
inflation falls, even self-defeating in amplifying them.

All such paths satisfy the private sector's transversality condition. Along
them, the conclusion is immediate, since the real interest rate converges to
$r$ and, with $\alpha >r$, real government debt converges to $b^{\ast }$. It
is worth establishing this result, however, by an argument that does not
rely on the convergence of inflation and real government debt, since the
same argument applies to the explosive inflationary paths considered below
and to the divergent debt paths of subsection 3.2. Let indeed%
\begin{equation}
q(t)\equiv e^{-\int_{0}^{t}[R(\pi (v))-\pi (v)]dv}b(t).  \label{q-def}
\end{equation}%
Differentiating (\ref{q-def}) with respect to time gives%
\begin{equation*}
\dot{q}(t)=e^{-\int_{0}^{t}[R(\pi (v))-\pi (v)]dv}\left\{ \dot{b}(t)-[R(\pi
(t))-\pi (t)]b(t)\right\} .
\end{equation*}%
Using (\ref{ODE-b}), $\dot{b}(t)-[R(\pi (t))-\pi (t)]b(t)=-\alpha b(t)-\bar{s}%
(t)$, so that%
\begin{equation}
\dot{q}(t)=-\alpha q(t)-e^{-\int_{0}^{t}[R(\pi (v))-\pi (v)]dv}\bar{s}(t).
\label{q-ode}
\end{equation}%
Solving (\ref{q-ode}) forward yields%
\begin{equation}
q(t)=e^{-\alpha t}q(0)-\int_{0}^{t}e^{-\alpha (t-v)}e^{-\int_{0}^{v}[R(\pi
(j))-\pi (j)]dj}\bar{s}(v)dv.  \label{q-solution}
\end{equation}%
The first term on the right-hand side of (\ref{q-solution}) vanishes as $%
t\rightarrow \infty $, since $\alpha >0$. As for the second term, along any path converging to $(y_{L},\pi
_{L}) $ the real interest rate converges to $R(\pi _{L})-\pi _{L}=r$
at an exponential rate, so that its deviation from $r$ is integrable and the
discount factor $e^{-\int_{0}^{v}[R(\pi (j))-\pi (j)]dj}$ is of order
$e^{-rv}$ up to a bounded multiplicative constant, and the discounted primary
surplus $e^{-\int_{0}^{v}[R(\pi (j))-\pi (j)]dj}\bar{s}(v)$
vanishes as $v\rightarrow \infty $, $\bar{s}(\cdot) $ being
bounded. Since the exponential kernel $e^{-\alpha (t-v)}$ integrates to at
most $1/\alpha $ and attaches vanishing weight to any bounded initial
interval of the integrand as $t\rightarrow \infty $, the second term
vanishes as well. Hence $q(t)\rightarrow 0$, so that the transversality
condition (\ref{TVC-equilibrium}) is satisfied along the entire trajectory,
regardless of its precise nonlinear shape.\footnote{%
The underlying reason is transparent from (\ref{q-ode}). The equilibrium
real interest rate enters the growth rate of real government debt and the
discount factor symmetrically, and therefore cancels out identically,
leaving the fiscal feedback coefficient $\alpha $ as the only force
governing the discounted value of liabilities, provided that the discounted
stream of exogenous primary surpluses vanishes.}

The pathology is all the more severe as the continuum of equilibria
converging to the liquidity trap includes trajectories originating
arbitrarily close to the intended steady state. Tracing the stable arm $%
Z_{L}Z_{L}$ backward in time, one finds that it spirals into $(y^{\ast
},\pi ^{\ast }) $, which, being a source, acts as an attractor when
time is reversed. There thus exists a heteroclinic connection running from
the target to the liquidity trap, as depicted in Figure 1. Trajectories
starting on it are initially indistinguishable from the intended
equilibrium, yet drift away from it in ever wider oscillations of output and
inflation until they settle onto the path leading to the trap. A formal proof
is provided in Appendix F: The argument is topological and does not depend
on the particular parameterization. Unlike in Benhabib, Schmitt-Groh\'{e}
and Uribe (2001a), where the saddle connection is one of two possible
outcomes---the other being convergence to a limit cycle around the
target---and is established for parameter values close to a bifurcation
point, here it is the only possible outcome, for all admissible parameter
values.\footnote{%
For preferences separable in consumption and real balances, Benhabib,
Schmitt-Groh\'{e} and Uribe (2001a, Proposition 1) show, for $r$ and $A-1$
positive and sufficiently small, that trajectories originating near the
target converge either to the liquidity trap, through a saddle connection,
or to a limit cycle, and illustrate the saddle connection numerically. In
the present model, the absence of closed orbits implied by the constant
divergence of the real block rules out the second alternative.}

What about the explosive inflationary paths emphasized by Cochrane (2011)?
Consider, for a given $b(0) $, an initial jump in inflation
above the target, off the stable arm $Z_{L}Z_{L}$. In the linearized model the implied trajectory diverges---with
oscillations of increasing amplitude in the case of complex eigenvalues, and
monotonically otherwise. Along a path on which inflation rose without bound,
the real interest rate would rise without bound as well, and such a path
would satisfy the private sector's transversality condition: The
discount factor would collapse to zero at a more than exponential rate, so
that the discounted primary surplus in (\ref{q-solution}) would vanish, and
$q(t) \rightarrow 0$. In the nonlinear economy, however, no such
path can be sustained. According to the dynamic IS curve (\ref{IS}), a real
interest rate permanently above the rate of time preference requires
consumption, and thus output, to grow; as output approaches the productive
capacity $\bar{y}$, hours approach the time endowment, the marginal
disutility of labor and real marginal cost explode, and the Phillips curve
(\ref{NKPC}) drives inflation down. Rather than rising without bound,
inflation thus turns down: The trajectory spirals outward around the
intended steady state---the absence of closed orbits precluding any periodic
equilibrium---until it falls below $\pi _{L}$, at which point inflation
keeps falling while the nominal interest rate approaches its zero lower
bound. The real interest rate then rises without bound again, and output
reaches the productive capacity in finite time: The trajectory ceases to
exist, and is not an equilibrium. Every trajectory that does not start on
the intended steady state or on $Z_{L}Z_{L}$ ends in the same way, in a
deflationary collapse below $\pi _{L}$ (see Appendix E). Explosive
inflationary paths are thus ruled out under an active stance of monetary
policy, but not by the private sector's asymptotic optimality conditions,
which they would satisfy: They are ruled out by the requirement that the
allocation they imply be feasible.

Taking stock, an active monetary/passive fiscal policy regime is affected,
from a global perspective, by real---not merely nominal---indeterminacy of
equilibrium, breaking down the stabilizing properties of the local analysis. The essence of the
global indeterminacy result hinges on the fact that fiscal policy behavior
ensures that the private sector's transversality condition is verified for
\textit{any} admissible time path of inflation. As a consequence, there are
no fiscal fundamentals that can act as an anchor for uniquely determining the
equilibrium: The only trajectories that are ruled out are those the economy
cannot produce.

\subsection{Passive Monetary/Active Fiscal Policy Regime}

A focal point of this subsection is to show that the pathology
characterizing an active monetary/passive fiscal policy regime---real
indeterminacy globally---still applies under a passive monetary/active
fiscal policy regime should the primary-surplus feedback reaction to
government liabilities remain positive, that is, $0<\alpha <r$.

To elucidate this point, and thereafter---in subsection 3.3---concentrate
the analysis on the polar policy configuration in which $\alpha =0$,
consider the case of the globally passive monetary policy reaction function
described in subsection 2.2, asymptotic to a straight line with slope $%
\varphi <1$ as $\pi \rightarrow \infty $, in conjunction with a fiscal
policy reaction function setting $0\leq \alpha <r$, thus generating \textit{%
per se} instability of real public debt. I also assume that the exogenous
component of the primary surplus is bounded away from zero, $\bar{s}(t) \geq \underline{s}>0$, so that the steady-state stock of real
government debt (\ref{SSb}) is positive.

Because monetary policy is globally passive, the $\dot{y}=0$-locus given by
the Fisher equation (\ref{pi-locus}) is horizontal only in correspondence of
the target inflation rate $\pi ^{\ast }$, and (\ref{y-dot}) now implies that
output rises below and falls above it. The $\dot{\pi}=0$-locus (\ref%
{y-locus}) is unchanged. The two loci intersect at the unique steady state $%
(y^{\ast },\pi ^{\ast }) $: Steady state uniqueness in
correspondence of the target inflation rate applies, dissolving the
existence of the unintended steady state prevailing in the active
monetary/passive fiscal policy regime.

\begin{figure}[!htb]
\centering
\includegraphics[width=0.78\textwidth]{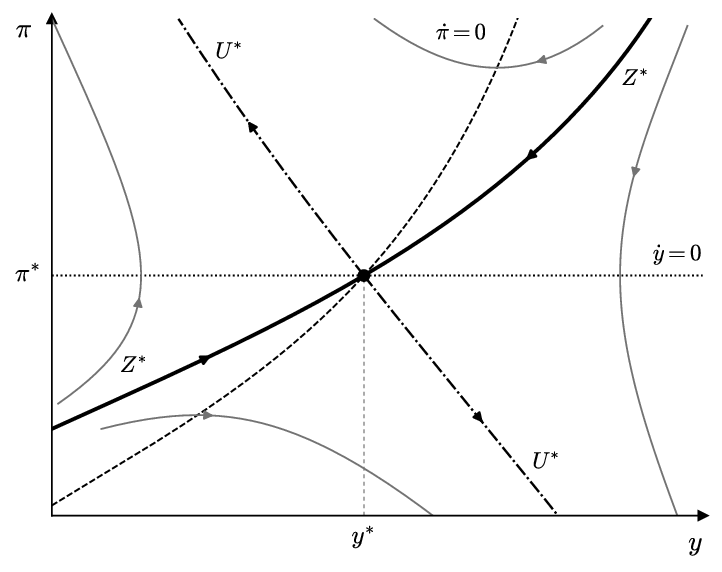}
\caption{Global Dynamics of the Real Block under a Passive Monetary/Active Fiscal Policy Regime}\medskip
\begin{minipage}{1.0\textwidth}
\footnotesize \textit{Notes}: The figure depicts the phase plane $(y,\pi) $ of the real
block. The dashed curve is the long-run Phillips curve $\dot{\pi}=0$; the
dotted line is the Fisher relation $\dot{y}=0$. The intended steady state is
a saddle, with stable arm $Z^{\ast }Z^{\ast }$ (solid) and unstable arm $%
U^{\ast }U^{\ast }$ (dash-dotted).
Off-saddle trajectories (grey) end either in a hyperinflation with
collapsing output or in a deflation with output approaching productive
capacity. The configuration is the same for any $\alpha \geq 0$. The figure
is drawn from a numerical illustration (see Appendix G).
\end{minipage}
\end{figure}

The foregoing properties lead to the global dynamics of the real block
characterized in Figure 2. Since $R^{\prime }(\pi ^{\ast }) <1$,
the Jacobian matrix (\ref{jacobian}) evaluated at $(y^{\ast },\pi
^{\ast }) $ has a negative determinant, and its eigenvalues are real
and of opposite sign. Let%
\begin{equation}
\mu \equiv \frac{r-\sqrt{r^{2}+4\kappa \frac{y^{\ast }}{\sigma (y^{\ast }) }\left[ 1-R^{\prime }(\pi ^{\ast }) \right] }}{2%
}<0  \label{mu}
\end{equation}%
denote the stable one. The real block exhibits a saddle point, with an
upward-sloping stable arm, $Z^{\ast }Z^{\ast }$, along which output and
inflation move together, and a downward-sloping unstable arm, $U^{\ast
}U^{\ast }$. Since the eigenvalue associated with real government debt, $%
r-\alpha $, is now positive, the full system has exactly one stable
eigenvalue, $\mu $, and the Blanchard-Kahn condition is still verified. In
particular, for a given $b(0) $, the only trajectory converging
asymptotically to $(y^{\ast },\pi ^{\ast },b^{\ast }) $ is now
given, to first order, by the saddle-path solution%
\begin{equation}
\pi (t) =\pi ^{\ast }+\frac{(r-\alpha) -\mu }{\left[
1-R^{\prime }(\pi ^{\ast }) \right] b^{\ast }}\left[ b(t) -b^{\ast }-\int_{t}^{\infty }e^{-(r-\alpha) (v-t) }\left[ \bar{s}(v) -\bar{s}^{\ast }\right] dv\right]
,  \label{ZZ1-PM/AF}
\end{equation}%
\begin{equation}
b(t) =b^{\ast }+\int_{t}^{\infty }e^{-(r-\alpha)
(v-t) }\left[ \bar{s}(v) -\bar{s}^{\ast }\right] dv
+e^{\mu t}\left[ b(0) -b^{\ast }-\int_{0}^{\infty }e^{-(r-\alpha) v}\left[ \bar{s}(v) -\bar{s}^{\ast }\right] dv%
\right] ,  \label{ZZ2-PM/AF}
\end{equation}%
\begin{equation}
y(t) =y^{\ast }+\frac{r-\mu }{\kappa }(\pi (t)
-\pi ^{\ast }) .  \label{ZZ3-PM/AF}
\end{equation}%
Inflation now reflects \textit{future} primary surpluses: The initial jump
in inflation responds to the initial real stock of government debt net of
the present discounted value of future deviations of the exogenous surplus
from its stationary value, so that news of lower future surpluses acts
exactly as a higher initial stock of debt.  Abstracting from such deviations, so that the direction of the initial
jump depends only on the initial real stock of government debt, for a given
$b(0) >b^{\ast }$, the passive stance of monetary policy enables
initial inflation $\pi (0) $ to jump up on the saddle path---above the target rate $\pi ^{\ast }$---in order to erode the
real value of government debt and rule out debt escalation, with output
rising above its steady-state level as the real interest rate falls. In this
sense, equilibrium inflation is dependent upon fiscal variables, in
accordance with the fiscal theory of the price level. According to (\ref%
{ZZ2-PM/AF}), and differently from equation (\ref{ZZ2-AM/PF})---where the
speed of convergence is governed by fiscal policy---real government debt
converges to the steady state at a velocity $-\mu $ governed by the central
bank's degree of accommodating behavior: From (\ref{mu}), $-\mu $ is
increasing in $1-R^{\prime }(\pi ^{\ast }) $.

The rate at which future surpluses are discounted in (\ref{ZZ1-PM/AF})
deserves attention. The saddle-path solution has the form of a valuation
equation for government debt, but the discount rate is $r-\alpha $, not the
real interest rate $r$. The Blanchard-Kahn selection amounts to requiring
that real government debt remain bounded---the stationarity requirement that
Cochrane (2023, p. 139) adopts explicitly as a constraint beyond the
transversality condition. For $0<\alpha <r$, this requirement is stronger
than the private sector's transversality condition, which discounts at the
real interest rate and, as shown below, is satisfied also along paths on
which real government debt grows without bound at the rate $r-\alpha $. The same discount rate appears in the constant-real-rate example of
Cochrane (2023, p. 142), where, as he notes, it has no particular
interpretation; here it is the discount rate implicit in the Blanchard-Kahn
selection. The two discount rates coincide only for $\alpha =0$.

Since the real block does not depend on fiscal variables, its stable arm $%
Z^{\ast }Z^{\ast }$ can be used to project the dynamics of the full system
onto the phase plane $(b,\pi) $, along which inflation evolves autonomously,
converging monotonically to the target. The resulting dynamics are depicted
in Figure 3 for a constant exogenous surplus, $\bar{s}(t) =\bar{%
s}^{\ast }$, which makes the system in $(b,\pi) $ autonomous;
with a time-varying surplus, the linearized dynamics are the same once $%
b(t) -b^{\ast }$ is replaced by its value net of the present
discounted value of future surplus deviations, as in (\ref{ZZ1-PM/AF}). The $%
\dot{b}=0$-locus, given by $b=\bar{s}^{\ast }/\left[ R(\pi)
-\pi -\alpha \right] $, is positively sloped, since the real
interest rate is decreasing in inflation, and $\dot{b}<(>) 0$ to
its left (right). The saddle path $ZZ$ is positively sloped as well,
and steeper than the $\dot{b}=0$-locus: From (\ref{ZZ1-PM/AF}), the two
slopes differ, around the steady state, by $-\mu /\{\left[ 1-R^{\prime
}(\pi ^{\ast }) \right] b^{\ast }\}>0$.

\begin{figure}[!htb]
\centering
\includegraphics[width=0.78\textwidth]{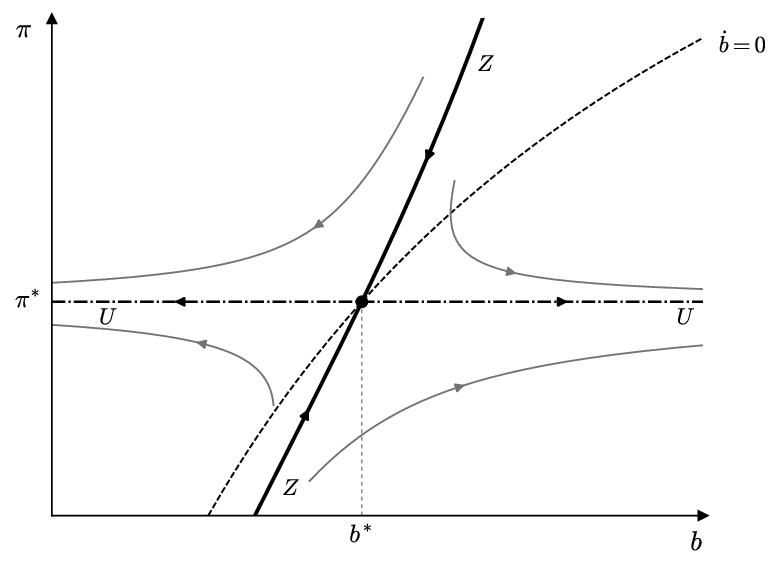}
\caption{Real Government Debt and Inflation under a Passive Monetary/Active Fiscal Policy Regime}\medskip
\begin{minipage}{1.0\textwidth}
\footnotesize \textit{Notes}: The figure depicts the phase plane $(b,\pi) $ along the
stable arm of the real block. The dashed curve is the $\dot{b}=0$-locus. The
saddle path $ZZ$ (solid) is steeper than the $\dot{b}=0$-locus; the
unstable arm $UU$ (dash-dotted) coincides globally with the
horizontal line $\pi =\pi ^{\ast }$. Off the saddle path (grey), inflation
converges to the target while real government debt diverges: Such
trajectories are equilibria for $0<\alpha <r$, but violate the
transversality or the no-Ponzi game condition for $\alpha =0$, in which case
the saddle path coincides with the government debt valuation equation. The
figure is drawn for $\alpha =0$ and a constant exogenous component of the
primary surplus, $\bar{s}(t) =\bar{s}^{\ast }$, from a
numerical illustration (see Appendix G); the configuration for $0<\alpha
<r$ is qualitatively identical.
\end{minipage}
\end{figure}

However, in the case in which the fiscal policy feedback parameter $\alpha $
continues to be positive---even though \textit{per se} insufficient to
stabilize real public debt according to (\ref{afp}) because $\alpha <r$---the
pathology that emerged in the active monetary/passive fiscal policy regime
remains operative here. To see this, two types of off-saddle trajectories
must be distinguished.

The first type comprises the trajectories that leave the stable arm $Z^{\ast }Z^{\ast }$ of
the real block. According to the dynamic system (\ref{IS})-(\ref{NKPC}), the
implied trajectory of $(y,\pi) $ curves asymptotically toward
the unstable arm $U^{\ast }U^{\ast }$, in one of two directions. Above the stable arm,
inflation embarks on an explosive path, rising at least exponentially, while
output collapses. Since the asymptotic slope of the monetary policy rule is
lower than unity, the real interest rate $R(\pi) -\pi $ falls
without bound, and the discount factor $e^{-\int_{0}^{v}[R(\pi (j))-\pi
(j)]dj}$ \textit{grows} at a more than exponential rate. The second term on
the right-hand side of (\ref{q-solution}) thus diverges to minus infinity,
for any $\alpha \geq 0$: Over any interval of fixed length $\delta >0$ ending
at $t$, the integrand is at least $e^{-\alpha \delta }\underline{s}$ times
the discount factor, which explodes. Hence $q(t) \rightarrow
-\infty $, and the no-Ponzi game condition (\ref{npg}) is violated: With a
real interest rate falling without bound, the present discounted value of the
primary surpluses that households would have to pay is infinite, and could
only be financed by a Ponzi scheme. Notice that it is not real government
debt that explodes along such a path---debt is in fact rapidly eroded by
inflation---but its discount factor. Below the stable arm, inflation falls
without bound, while the nominal interest rate approaches its zero lower
bound; the real interest rate rises without bound, and, as in subsection
3.1, output reaches the productive capacity in finite time, so that the
trajectory ceases to exist. Appendix D shows that every trajectory of the
real block other than those on $Z^{\ast }Z^{\ast }$ ends in one of these two ways.
Hyperinflations are thus ruled out by the private sector's asymptotic
conditions, deflations by the economy's productive capacity.

The second type comprises the trajectories that remain on the stable arm $Z^{\ast }Z^{\ast }$
of the real block, but start off the saddle path $ZZ$ in the phase
plane $(b,\pi) $. Consider for a given value $b(0)$ an initial
jump in inflation below the saddle path. According to Figure 3, the implied trajectory of $(b,\pi) $ curves asymptotically toward the unstable arm of the saddle path, $%
UU$---which, unlike the unstable arm $U^{\ast }U^{\ast }$ of the real
block, calls for no
local approximation, since it coincides both locally and globally with the
horizontal line%
\begin{equation*}
\pi =\pi ^{\ast }.
\end{equation*}%
Inflation and output gradually revert to their steady-state levels, while
real government debt embarks on an unstable path. Along $UU$, the
real interest rate settles at its steady-state value, $R(\pi ^{\ast
}) -\pi ^{\ast }=r$, and the law of motion (\ref{ODE-b}) becomes%
\begin{equation*}
\dot{b}=(r-\alpha) b-\bar{s},
\end{equation*}%
so that $b(t) $ grows without bound at the asymptotic rate $%
r-\alpha >0$, the primary surplus failing to respond to debt accumulation
strongly enough to stabilize it.

The key point is that, regardless of the exact nonlinear shape of the
trajectory, such a path necessarily satisfies the private sector's
transversality condition. Let indeed $q(t) $ be defined as in (%
\ref{q-def}). The derivation of (\ref{q-ode}) makes use of (\ref{ODE-b})
alone, which is common to both policy configurations, so that (\ref%
{q-solution}) continues to hold unchanged. The first term on its right-hand
side vanishes as $t\rightarrow \infty $, since $\alpha >0$. As for the
second term, since inflation converges to the target at the exponential rate
$\mu $, the deviation of the real interest rate from $r$ is integrable, so
that the discount factor $e^{-\int_{0}^{v}[R(\pi (j))-\pi (j)]dj}$ remains
of order $e^{-rv}$ up to a bounded multiplicative constant. The discounted
primary surplus $e^{-\int_{0}^{v}[R(\pi (j))-\pi (j)]dj}\bar{s}(v)$ thus
vanishes as $v\rightarrow \infty $, $\bar{s}(\cdot) $ being
bounded. Since the exponential kernel $e^{-\alpha (t-v)}$ integrates to at
most $1/\alpha $ and attaches vanishing weight to any bounded initial
interval of the integrand as $t\rightarrow \infty $, the second term
vanishes as well. Hence $q(t)\rightarrow 0$, so that the transversality
condition (\ref{TVC-equilibrium}) is satisfied exactly along the entire
trajectory, regardless of its precise nonlinear shape. The same conclusion
holds for initial jumps above the saddle path, along which real government debt explodes in the opposite
direction: The second term in (\ref{q-solution}) is unaffected, depending
only on the path of inflation, while the first vanishes irrespective of the
sign of $q(0)$.

Along all such trajectories, the government debt valuation equation---which,
by integrating the government's flow budget constraint (\ref{gibc}) forward,
is equivalent to the transversality condition---holds as an
identity---whatever the initial inflation rate on the stable arm of the real
block---with future primary surpluses $s=\bar{s}+\alpha b$ adjusting
endogenously to the debt path they themselves generate. It therefore
restricts nothing. This is the nonlinear counterpart, in a New Keynesian
economy with sticky prices and an endogenous real interest rate, of the
observation in Cochrane (2023, Sec. 5.4.2) that, under a constant real
interest rate, the valuation equation then holds for any price level. As a
main extension of the above results, there exists a continuum of initial
inflation rates, both above and below the saddle path, that can be supported
as equilibrium outcomes. Global indeterminacy stands out again.

\subsection{Pure Fiscal Dominance}

I have now the necessary elements to demonstrate the paper's central
argument: The pathology of global indeterminacy is eradicated---restoring
full equilibrium uniqueness---as soon as one passes to a regime of pure
fiscal dominance whereby fiscal policy is globally (not merely locally)
non-Ricardian.

To show this point, consider the polar policy configuration in which $\alpha
=0$, maintaining the globally passive monetary policy reaction function and
the assumption $\bar{s}(t) \geq \underline{s}>0$ of subsection
3.2. Imposing $\alpha =0$ in equation (\ref{q-solution}) and observing from (%
\ref{q-def}) that $q(0) \equiv b(0) $ yields%
\begin{equation*}
q(t)=b(0)-\int_{0}^{t}e^{-\int_{0}^{v}[R(\pi (j))-\pi (j)]dj}\bar{s}(v)dv.
\end{equation*}%
Requiring that the private sector's transversality condition $\underset{%
t\rightarrow \infty }{\lim }q(t) =0$ hold now introduces a
\textit{restriction} on the equilibrium path of inflation, given by%
\begin{equation}
b(0)=\frac{B_{0}}{P_{0}}=\int_{0}^{\infty }e^{-\int_{0}^{t}[R(\pi (j))-\pi
(j)]dj}\bar{s}(t)dt.  \label{gdve}
\end{equation}%
Equation (\ref{gdve}) is the government debt valuation equation \`{a} la
Cochrane (2011, 2023): It asserts that in equilibrium the initial real stock
of government debt must equal the present discounted value of real primary
surpluses. Since prices are sticky, its left-hand side is predetermined.
As in Cochrane (2023, Ch. 5), the discount factor $%
e^{-\int_{0}^{t}[R(\pi (j))-\pi (j)]dj}$ is endogenous; differently from his
linearized analysis, it enters here in its exact nonlinear form: The
right-hand side depends on the entire equilibrium path of inflation, through
the real interest rate at which future primary surpluses are discounted.

Consider first the trajectories of the first type, those leaving the stable
arm $Z^{\ast }Z^{\ast }$ of the real block. The arguments of subsection 3.2 apply unchanged,
since they hold for any $\alpha \geq 0$. Along hyperinflationary
trajectories, the right-hand side of (\ref{gdve}) is infinite, and the
valuation equation cannot hold for any finite initial stock of debt: The
endogenous discount factor, by exploding as the real interest rate falls
without bound, rules out self-fulfilling hyperinflations. Deflationary trajectories, as shown in subsection 3.2, cease to exist in
finite time, as output reaches the productive capacity of the economy.

Consider next the trajectories that remain on the stable arm $Z^{\ast }Z^{\ast }$ of the
real block. Since output and inflation converge to the steady state along
it, each initial inflation rate $\pi (0) $ on $Z^{\ast }Z^{\ast }$ uniquely determines the
entire path of inflation, and thus the right-hand side of (\ref{gdve}).
Denote it by $V(\pi (0)) $, so that the equilibrium requires%
\begin{equation}
b(0)=V(\pi (0)) .  \label{gdve-V}
\end{equation}%
The properties of $V(\cdot) $ follow directly from those of the
stable arm and of the monetary policy reaction function. Along $Z^{\ast }Z^{\ast }$,
inflation converges monotonically to the target (see Appendix D), so that
trajectories starting from a higher initial inflation rate exhibit higher
inflation at every date. With the central bank accommodating rather than
fighting such inflation, the equilibrium real interest rate is lower at
every date, and the discount factor higher at every horizon. Since primary
surpluses are positive, it follows that $V(\cdot) $ is strictly
increasing: Higher initial inflation, by lowering real interest rates along
the entire adjustment path, raises the present discounted value of the
primary surpluses that back government debt. A strictly increasing function
takes each value at most once. The government debt valuation equation (\ref%
{gdve-V}) therefore admits at most one solution, and exactly one whenever
the initial real stock of government debt lies in the range of $V(\cdot) $, an interval that contains the steady-state stock $b^{\ast }$
in its interior.\footnote{%
In the numerical illustration underlying the figures, the range of $V(\cdot) $ extends from values close to zero to about one and a half
times $b^{\ast }$. Its upper bound reflects the productive capacity of the
economy: Along the upper branch of the stable arm, output approaches $\bar{y}$%
, and further increases in initial inflation can only be sustained for a
vanishing interval of time, which limits the fall in real interest rates
that inflation can engineer. For a larger initial stock of debt, no
equilibrium exists.} Uniqueness thus requires no further assumption: Since
the initial price level is predetermined, endogenous discounting operates
only on the right-hand side of the valuation equation, and it does so
monotonically. The positive local
slope of the saddle path, from (\ref{ZZ1-PM/AF}), would not suffice: It
guarantees that $V(\cdot) $ is increasing only near the steady
state, and it is the global monotonicity of $V(\cdot) $ that
rules out a folding of the saddle path, and hence multiple initial inflation
rates for a given $b(0)$.

It is in this precise sense that, under pure fiscal dominance, the
government debt valuation equation \textit{is} the saddle path. In the phase
plane $(b,\pi) $ of Figure 3, the saddle path $ZZ$ is
the locus of initial conditions from which real government debt converges to
its steady state; with $\alpha =0$, it coincides with the graph of the
function $V(\cdot) $. Off the saddle path, by contrast, the
transversality condition fails. Indeed, along any trajectory starting on the
stable arm of the real block,%
\begin{equation}
\underset{t\rightarrow \infty }{\lim }q(t) =b(0)-V(\pi (0)) ,  \label{q-limit-alfa=0}
\end{equation}%
the present discounted value on the right-hand side being evaluated along
the trajectory actually followed. The discounting term in (\ref{q-solution})
no longer vanishes, as it did for any $\alpha >0$: The fiscal feedback
coefficient that previously discounted away the initial stock of debt at the
rate $e^{-\alpha t}$ is now absent, and $\lim_{t\rightarrow \infty }q(t) $ equals the gap between the initial stock of real government debt
and the discounted backing that the trajectory itself generates.

Consider first, as depicted in Figure 3, a trajectory originating below the
saddle path, for which $b(0)>V(\pi (0)) $. Inflation converges
to the target rate along the unstable arm $UU$, whereas real
government debt diverges without bound, growing asymptotically at the
steady-state real interest rate. Hence $\lim_{t\rightarrow \infty }q(t) >0$, and the transversality condition (\ref{TVC-equilibrium}) is
violated. The household could raise lifetime utility by consuming part of
the wealth it is instead accumulating without limit; such a trajectory,
though consistent with the laws of motion of the economy and with the
borrowing limit, is not an optimal plan, and therefore not an equilibrium.

Consider now a trajectory originating above the saddle path, for which $%
b(0)<V(\pi (0)) $. Inflation again converges to the target
rate, but real government debt now diverges in the opposite direction, so
that $\lim_{t\rightarrow \infty }q(t) <0$. Here the violation is
more radical still: The no-Ponzi game condition (\ref{npg}) fails, since the
present discounted value of the household's net asset position becomes
negative in the limit. Real government debt turning negative means that
households become net debtors of the government, borrowing from it to pay
the primary surpluses that are no longer needed to service any debt, and
rolling over the resulting liabilities forever. No constraint on the
government is violated: Its policy remains well defined along the
trajectory. Such a trajectory does not merely fail to be optimal: It lies
outside the set of plans available to the household.

Off-saddle trajectories are thus eliminated from all sides, and for reasons
of different strength: Below the saddle path by the failure of optimality,
above it and along hyperinflations by the failure of feasibility of the
household's plan, along deflations by the failure of feasibility of the
allocation itself. What remains is the saddle path alone, and on it the
single point pinned down by the government debt valuation equation. Inflation
jumps to the corresponding point on the saddle path, output jumps
accordingly along the stable arm of the real block, and the economy follows
the unique trajectory converging to $(y^{\ast },\pi ^{\ast },b^{\ast
}) $.

The full determination of the perfect-foresight equilibrium in the
neighborhood around the steady state can be obtained in closed form by
linearizing the government debt valuation equation (\ref{gdve}) around the
steady state, where $b^{\ast }=\bar{s}^{\ast }/r$. This also provides a local
check of the global equivalence just established. Linearizing the real
interest rate gives $R(\pi) -\pi \simeq r-\left[ 1-R^{\prime
}(\pi ^{\ast }) \right] (\pi -\pi ^{\ast }) $, so
that, using $\pi (j) -\pi ^{\ast }=(\pi (0) -\pi ^{\ast
}) e^{\mu j}$ along the stable arm, the discount factor is, to first
order,%
\begin{equation*}
e^{-\int_{0}^{t}[R(\pi (j))-\pi (j)]dj}\simeq e^{-rt}\left[ 1+\left[
1-R^{\prime }(\pi ^{\ast }) \right] \frac{1-e^{\mu t}}{-\mu }%
(\pi (0) -\pi ^{\ast }) \right] .
\end{equation*}%
Substituting into (\ref{gdve}), with $\bar{s}(t) =\bar{s}^{\ast
}+\left[ \bar{s}(t) -\bar{s}^{\ast }\right] $, and neglecting
second-order terms yields%
\begin{equation}
b(0)=b^{\ast }+\int_{0}^{\infty }e^{-rt}\left[ \bar{s}(t) -\bar{%
s}^{\ast }\right] dt+\frac{1-R^{\prime }(\pi ^{\ast }) }{r-\mu }%
b^{\ast }(\pi (0) -\pi ^{\ast }) .  \label{gdve-lin}
\end{equation}%
Since $b(0)$ is predetermined, (\ref{gdve-lin}) is an equation in the single
unknown $\pi (0) $, whose solution is%
\begin{equation}
\pi (0) =\pi ^{\ast }+\frac{r-\mu }{\left[ 1-R^{\prime }(\pi ^{\ast
}) \right] b^{\ast }}\left[ b(0)-b^{\ast }-\int_{0}^{\infty }e^{-rt}%
\left[ \bar{s}(t) -\bar{s}^{\ast }\right] dt\right] ,
\label{pi0-closed}
\end{equation}%
and, from (\ref{ZZ3-PM/AF}), the initial level of output is%
\begin{equation}
y(0) =y^{\ast }+\frac{r-\mu }{\kappa }(\pi (0) -\pi ^{\ast }) .
\label{y0-closed}
\end{equation}%
Equation (\ref{pi0-closed}) is precisely the saddle-path solution (\ref%
{ZZ1-PM/AF}) evaluated at $\alpha =0$ and $t=0$: The linearized valuation
equation and the linearized saddle path are one and the same relation, the
local counterpart of the global equivalence established above. For $\alpha
>0$, by contrast, the two relations no longer coincide. The
saddle-path solution (\ref{ZZ1-PM/AF}) still has the form of a valuation
equation, but one in which future surpluses are discounted at the rate $%
r-\alpha $ rather than at the real interest rate $r$: It is not implied by
the transversality condition, which, as shown in subsection 3.2, holds along
every trajectory of the second type. Only under pure fiscal dominance do the
two discount rates, and hence the two relations, coincide.

Initial inflation exceeds the target if and only if the initial real stock
of government debt exceeds the present discounted value of exogenous primary
surpluses evaluated at the steady-state real interest rate, $b^{\ast
}+\int_{0}^{\infty }e^{-rt}\left[ \bar{s}(t) -\bar{s}^{\ast }%
\right] dt=\int_{0}^{\infty }e^{-rt}\bar{s}(t) dt$. The gap is
closed by a fall in the real interest rate along the entire adjustment path,
which raises the discounted value of the surplus stream: Equilibrium
inflation is a fiscal phenomenon, shaped by the speed of convergence $-\mu $
and by the degree of accommodation embedded in the central bank's
interest-rate policy, $1-R^{\prime }(\pi ^{\ast }) $. News of
lower future surpluses raises initial inflation exactly as a higher initial
stock of debt would.

It is worth emphasizing what this argument does not require. No contingent
threat is invoked, no announced regime switch, no commitment to a policy
that the government would not wish to carry out. The monetary-fiscal
configuration is time-invariant throughout, and off-saddle trajectories are
ruled out not because the government promises to \textquotedblleft
punish\textquotedblright\ them but because the private sector, optimizing
under an initial endowment of nominal wealth and predetermined prices, would
never choose them, or could not carry them out. It is in this precise sense
that pure fiscal dominance provides the nominal anchor: Not by threatening
to \textquotedblleft blow up the economy\textquotedblright\ should
expectations depart from the intended equilibrium, but by leaving a single
initial configuration compatible with the household's asymptotic conditions
and with the economy's resources---the one at which the government debt
valuation equation holds. Any revision of expectations away from it would
displace the economy onto a trajectory violating either the transversality
condition or the no-Ponzi game condition, or onto an allocation that cannot
be produced, and would therefore not be self-fulfilling.

\section{Discussion and Conclusion}

As aptly emphasized by Cochrane (2011, p. 587), \textquotedblleft \lbrack
t]he transversality condition is a consumer first-order
condition\textquotedblright\ for the dynamic optimal plan. It is hard to
justify disregarding it, as models relying only on local dynamics implicitly
do, or replacing it with the requirement that equilibrium paths remain
bounded.
The local classification of monetary-fiscal regimes, in other words,
obscures a pathology that only a global analysis can highlight: It correctly
signs the local eigenvalues of the dynamical system, but abstracts from
whether trajectories that fail the local convergence criterion may
nonetheless satisfy the economy's asymptotic equilibrium conditions, once
nonlinear dynamics are fully accounted for. From this perspective, the
results shown in this paper help highlight a general point: The
Blanchard-Kahn conditions for uniqueness and stability of equilibrium
maintain a general validity if and only if off-saddle trajectories under the
monetary-fiscal mix at hand---not under a contingent policy
\textquotedblleft threat to blow up the economy\textquotedblright\
(Cochrane, 2011, p. 567)---can be ruled out because they violate either the
private sector's transversality condition or its no-Ponzi game condition, or
because the allocations they imply cannot be produced. Should this criterion
not be satisfied, it is hard to contend that the dynamic general equilibrium
model under study is really \textit{optimizing}, and, most importantly, the
\textit{equilibrium} is fatally affected by global indeterminacy.

Reconsidering the Leeper dichotomy through the lens of a global analysis,
global determinacy does not hold whenever fiscal policy remains globally
Ricardian---as is the case not only under the canonical active
monetary/passive fiscal regime, but also under a passive monetary/active
fiscal regime whenever the response of primary surpluses to government
liabilities remains strictly positive, however small. In both cases fiscal
policy guarantees that the discounted value of government liabilities
vanishes along any path the economy may follow, so that the transversality
condition is slack and cannot discriminate among equilibria. What differs
across the two regimes is only the form taken by the indeterminacy: a
continuum of decelerating paths converging to a liquidity trap in the
former, a continuum of paths along which inflation returns to the target in
the latter. Global non-Ricardianness, with
primary surpluses set independently of the real value of government
liabilities, is the essential element for obtaining global determinacy. This
is consistent with the view in Cochrane (2023, Secs. 5.4.1--5.4.3) that an
active fiscal policy can respond to debt, provided that primary surpluses
respond to debt valued at a target price level, and not to the revaluation
of government liabilities induced by the equilibrium price level itself:
Such a response makes primary surpluses independent of the equilibrium path,
and thus gives them the nature of an exogenous surplus stream, as under the
regime of pure fiscal dominance considered here.

The analysis carries three implications that bear directly on the
fiscal theory of the price level as developed by Cochrane (2023). The first
concerns the government debt valuation equation itself. Under pure fiscal
dominance, the valuation equation \textit{is} the saddle path: The locus of
initial conditions from which real government debt converges to its steady
state coincides with the locus of initial conditions satisfying the
transversality condition. Under a positive fiscal feedback, by contrast, the
valuation equation holds along every candidate path---future primary
surpluses adjusting endogenously to the debt path they generate---and
therefore restricts nothing, while the Blanchard-Kahn saddle path is only one
among a continuum of equilibria. The linearized solution makes the
difference visible in a single parameter: The saddle path discounts future
primary surpluses at the rate $r-\alpha $, the transversality condition at
the real interest rate $r$. The equivalence between the saddle-path
apparatus and the fiscal theory, in other words, is not general: It holds
exactly, and only, under pure fiscal dominance, when the two discount rates
coincide.

The second implication concerns endogenous discounting. With sticky prices,
the valuation equation determines initial inflation rather than the initial
price level, and it does so through the real interest rate at which future
primary surpluses are discounted. Endogenous discounting is not a nuisance to
be linearized away. It is precisely what rules out self-fulfilling
hyperinflations under pure fiscal dominance---the real interest rate falling
without bound and the present discounted value of primary surpluses
diverging---and it is what delivers uniqueness, since higher initial
inflation, by lowering real interest rates along the entire adjustment path,
raises the discounted backing of government debt monotonically. The
resulting equilibrium inflation rate depends on the gap between initial real
debt and the present value of exogenous primary surpluses at the
steady-state real interest rate---so that news of lower future surpluses
raises initial inflation exactly as higher initial debt does---scaled by the
speed of convergence and by the degree of accommodation embedded in monetary
policy: When globally
determinate, inflation is a fiscal phenomenon, shaped by the central bank's
interest-rate policy.

The third implication concerns the explosive inflationary paths that,
according to Cochrane (2011), the private sector's asymptotic conditions
cannot rule out under an active stance of monetary policy. As the present
analysis shows, the transversality condition alone does not exclude them,
for along such paths the real interest rate explodes and government
liabilities are discounted away. In the nonlinear New Keynesian economy,
however, they do not exist: An ever-rising real interest rate requires
consumption to grow without bound, which the economy cannot produce. The
explosive oscillations of the linearized model end, globally, either in the
liquidity trap or in a deflationary collapse that is not an equilibrium
either. Excluding explosive inflation, however, does not rescue the Taylor
principle: The continuum of decelerating paths leading to the liquidity trap,
including paths originating arbitrarily close to the target, survives intact.
And what rules out explosive paths in the active regime---the economy's
productive capacity---differs in nature from what rules them out under pure
fiscal dominance---the private sector's no-Ponzi game condition.

The results also suggest a note of caution for the empirical assessment of
monetary-fiscal regimes. Estimated fiscal reaction functions typically
document a positive response of primary surpluses to government debt (Bohn,
1998), which the literature on fiscal sustainability reads as evidence that
the government's intertemporal budget constraint is satisfied. Leeper and Li
(2017) show that such estimates are subject to simultaneity bias and cannot
by themselves identify whether fiscal policy is active or passive. The
present analysis points to a complementary, theoretical limitation: Even a
correctly identified positive response of primary surpluses to the real value
of government liabilities, however small, implies that fiscal policy is
globally Ricardian, and thus unable to provide the nominal anchor, unless it
is exactly zero. As Cochrane (2023, Sec. 5.4.1) emphasizes, however, a
positive regression coefficient of surpluses on debt does not by itself
establish such a response: Surpluses responding to debt valued at a target
price level, and not to its revaluation by inflation, would deliver the same
equilibrium correlation.

The paper's argument requires no contingent switch in policy, which,
according to Cochrane (2011), would imply a questionable game between the
government and atomistic agents. The monetary-fiscal configuration is
time-invariant, and off-saddle trajectories are excluded not because the
government threatens to punish them, but because they are inconsistent with
private optimization or with the economy's resources under that very
configuration. By contrast, the requirement of a stationary debt-to-GDP
ratio (Cochrane, 2023, p. 139), which restores uniqueness under a positive
fiscal feedback, is a constraint beyond the private sector's optimality
conditions: Under a given, time-invariant fiscal rule, it acts as a
criterion of equilibrium selection rather than as an implication of
optimizing behavior. Global determinacy is, in this sense, a matter of
consistency between a non-state-contingent policy regime and the optimizing
behavior of forward-looking agents.

Several extensions are left for future research. Calvo pricing introduces
price dispersion as an additional state variable and would require a global
analysis in higher dimension. Money-in-the-utility-function preferences
nonseparable in consumption and real balances would make the divergence of
the real block vary along trajectories, so that periodic equilibria may
arise under active monetary policy (Benhabib, Schmitt-Groh\'{e} and Uribe,
2001a); under passive monetary policy, the unique steady state of the real
block is a saddle, which no closed orbit can encircle, but the remaining
steps of the global analysis would need to be re-established. Long-term government debt, which Cochrane
(2023) shows to be central to the dynamics of fiscal inflation, would add a
further channel through which the endogenous discount factor shapes the
valuation of government liabilities. And stochastic fluctuations in primary
surpluses would raise the question of how global determinacy interacts with
the propagation of fiscal shocks.

In conclusion, global dynamics are not a technical refinement of monetary
theory: They decide which policy regimes deliver a determinate equilibrium.
Local determinacy, whether obtained through the Taylor principle or through
an active fiscal stance, is a property of the neighborhood of the target; it
says nothing about the trajectories along which the economy may be led away
from it. Once those trajectories are admitted into the analysis, only one
configuration survives: a fiscal authority that sets primary surpluses
independently of the real value of government liabilities. Only then is every out-of-fundamentals
path inconsistent with private optimization or with the economy's resources,
and only then does the saddle path---singled out by the government debt
valuation equation---become \textit{the} equilibrium.

\appendix

\section*{Appendix A. The Nonlinear Phillips Curve}

Let $\lambda $, $\mu _{P}$, and $\nu $ denote the multipliers associated
with the budget constraint (\ref{hbc}), the law of motion $\dot{P}_{j}=\pi
_{j}P_{j}$, and the demand constraint $f(h_{j}) =(P_{j}/P) ^{\eta }y$, respectively. The current-value Hamiltonian of
household-firm $j$ is%
\begin{multline*}
H=u(c) -z(h_{j}) -\frac{\gamma }{2}(\pi
_{j}-\pi ^{\ast }) ^{2}+\lambda \left[ (R-\pi) b+\frac{%
P_{j}}{P}f(h_{j}) -c-\tau \right] \\
+\mu _{P}\pi _{j}P_{j}+\nu \left[ f(h_{j}) -\left( \frac{P_{j}}{P%
}\right) ^{\eta }y\right] .
\end{multline*}%
The first-order conditions with respect to $c$, $h_{j}$, and $\pi _{j}$ are $%
u^{\prime }(c) =\lambda $, $z^{\prime }(h_{j})
=(\lambda P_{j}/P+\nu) f^{\prime }(h_{j}) $, and $%
\gamma (\pi _{j}-\pi ^{\ast }) =\mu _{P}P_{j}$; the co-state
equations are (\ref{cs10}) and $\dot{\mu}_{P}=r\mu _{P}-\partial H/\partial
P_{j}$, with $\partial H/\partial P_{j}=\lambda f(h_{j}) /P+\mu
_{P}\pi _{j}-\nu \eta (P_{j}/P) ^{\eta -1}y/P$. In a symmetric
equilibrium, $P_{j}=P$ and $f(h) =y$, so that $\nu =z^{\prime
}/f^{\prime }-\lambda $. Letting $x\equiv \mu _{P}P=\gamma (\pi -\pi
^{\ast }) $ and using $\dot{x}=\dot{\mu}_{P}P+x\pi $, one obtains%
\begin{equation*}
\dot{x}=rx-y\left[ (1+\eta) \lambda -\eta \frac{z^{\prime }}{%
f^{\prime }}\right] ,
\end{equation*}%
which, using $\lambda =u^{\prime }(y) $, gives (\ref{NKPC})-(\ref%
{Gamma}).

\section*{Appendix B. Properties of $\Gamma (\cdot) $}

Write $\gamma \Gamma (y) =-\left\vert 1+\eta \right\vert
yu^{\prime }(y) +\left\vert \eta \right\vert yz^{\prime }(h) /f^{\prime }(h) $, with $h=f^{-1}(y) $. The
second term is strictly increasing in $y$, since $z^{\prime }$ is increasing
and $f^{\prime }$ nonincreasing. As for the first, $d\left[ yu^{\prime
}(y) \right] /dy=u^{\prime }(y) \left[ 1-\sigma
(y) \right] \leq 0$. Hence $\Gamma (\cdot) $ is
strictly increasing. As $y\rightarrow \bar{y}$, $h\rightarrow \bar{h}$ and $%
z^{\prime }(h) \rightarrow \infty $, so that $\Gamma (y) \rightarrow \infty $. As $y\rightarrow 0$, $yz^{\prime }(h) /f^{\prime }(h) =z^{\prime }(h) f(h) /f^{\prime }(h) \rightarrow 0$, while $yu^{\prime
}(y) $, being nonincreasing, converges to a limit in $\left(
0,\infty \right] $; hence $\Gamma (y) \rightarrow \Gamma
_{0}\equiv -\left( \left\vert 1+\eta \right\vert /\gamma \right)
\lim_{y\rightarrow 0}yu^{\prime }(y) <0$. By continuity and
strict monotonicity, $\Gamma (y) =0$ has a unique solution $%
y^{\ast }\in (0,\bar{y}) $.

\section*{Appendix C. Absence of Closed Orbits}

Let $\theta \equiv \ln u^{\prime }(y) $, a strictly decreasing
function of $y$, with $\theta >\theta _{\min }\equiv \ln u^{\prime }(\bar{y}) $. From (\ref{uc})-(\ref{cs10}) and (\ref{NKPC}), the real
block can be written as%
\begin{equation*}
\dot{\theta}=r+\pi -R(\pi) ,\qquad \dot{\pi}=r(\pi -\pi
^{\ast }) -\Gamma (y(\theta)) .
\end{equation*}%
Since $\dot{\theta}$ does not depend on $\theta $ and $\dot{\pi}$ depends on
$\theta $ only through $\Gamma $, the divergence of the vector field is $%
\partial \dot{\theta}/\partial \theta +\partial \dot{\pi}/\partial \pi =r>0$
everywhere. By Bendixson's criterion (e.g., Perko, 2001), there
exists no closed orbit. More generally, there exists no closed curve formed
by trajectories and steady states (homoclinic loops or heteroclinic cycles):
The flow of the vector field across such a curve is zero, whereas, by the
divergence theorem, it should equal $r$ times the area of the region it
encloses. All the arguments below are conducted in the coordinates $(\theta ,\pi) $ and translated into $(y,\pi) $ through the
monotone change of variable $y=(u^{\prime }) ^{-1}\left(
e^{\theta }\right) $.

\section*{Appendix D. Passive Monetary Policy}

The passive case is treated first, since the invariance argument developed
here is used again for the active case in Appendix E. Under the globally
passive rule, $R(\pi) -\pi $ is strictly
decreasing and equals $r$ at $\pi ^{\ast }$, so that $\dot{y}<0$ for $\pi
>\pi ^{\ast }$ and $\dot{y}>0$ for $\pi <\pi ^{\ast }$. Along any
trajectory, $d\dot{\pi}/dt=r\dot{\pi}-\Gamma ^{\prime }(y) \dot{y%
}$. Consider the regions $U\equiv \left\{ \pi >\pi ^{\ast },\dot{\pi}%
>0\right\} $, $D\equiv \left\{ \pi <\pi ^{\ast },\dot{\pi}<0\right\} $, $%
A\equiv \left\{ \pi >\pi ^{\ast },\dot{\pi}\leq 0\right\} $, and $B\equiv
\left\{ \pi <\pi ^{\ast },\dot{\pi}\geq 0\right\} $.

\textit{Classification of trajectories.} The region $U$ is positively
invariant: On its boundary $\dot{\pi}=0$, with $\pi >\pi ^{\ast }$, one has $%
d\dot{\pi}/dt=-\Gamma ^{\prime }(y) \dot{y}>0$, and $\pi $ is
increasing within $U$. The same argument shows that $D$ is positively
invariant. In $A$, $\Gamma (y) \geq r(\pi -\pi ^{\ast
}) >0$, so that $y>y^{\ast }$; output is decreasing and inflation
nonincreasing. A trajectory in $A$ either reaches the boundary $\dot{\pi}=0$
and enters $U$, or reaches the line $\pi =\pi ^{\ast }$, where $\dot{\pi}%
=-\Gamma (y) <0$, and enters $D$, or remains in $A$ forever;
in the latter case $(y,\pi) $ is monotone and bounded, hence
convergent, and its limit is the steady state. The region $B$ is treated
symmetrically. Hence every trajectory either converges to $(y^{\ast
},\pi ^{\ast }) $ or enters $U$ or $D$. The two branches of the stable
arm lie in $A$ (with $y>y^{\ast }$, $\pi >\pi ^{\ast }$) and in $B$ (with $%
y<y^{\ast }$, $\pi <\pi ^{\ast }$), and both output and inflation are
monotone along each of them.

\textit{Hyperinflations violate the no-Ponzi game condition.} In $U$, output
is decreasing, so that $\Gamma (y(t)) \leq \Gamma
(y(t_{0})) $ and $\dot{\pi}\geq r(\pi -\pi
^{\ast }) -\Gamma (y(t_{0})) $, with $\dot{\pi%
}(t_{0}) >0$. By comparison, $\pi $ diverges at least
exponentially. Since $R(\cdot) $ is asymptotic to a straight
line with slope $\varphi <1$, the real interest rate $R(\pi)
-\pi $ diverges to $-\infty $: For every $M>0$ there exists $T_{M}$ such that
$R(\pi (t)) -\pi (t) \leq -M$ for $t\geq
T_{M}$. The discount factor $\Delta (t) \equiv
e^{-\int_{0}^{t}[R(\pi (v))-\pi (v)]dv}$ is then increasing for $t\geq T_{M}$
and grows faster than any exponential. Since the integrand in (\ref%
{q-solution}) is positive, for any $\delta >0$ and $t\geq T_{M}+\delta $,%
\begin{equation*}
q(t) \leq e^{-\alpha t}b(0) -\int_{t-\delta
}^{t}e^{-\alpha (t-v) }\Delta (v) \bar{s}(v) dv\leq e^{-\alpha t}b(0) -\delta e^{-\alpha \delta }%
\underline{s}\Delta (t-\delta) \rightarrow -\infty ,
\end{equation*}%
for any $\alpha \geq 0$, which violates (\ref{npg}).

\textit{Deflations are not solutions.} In $D$, output is increasing, so that
$\Gamma (y(t)) \geq \Gamma (y(t_{0})) $ and $\dot{\pi}\leq r(\pi -\pi ^{\ast })
-\Gamma (y(t_{0})) $, with $\dot{\pi}(t_{0}) <0$: As long as the solution exists, inflation diverges to $%
-\infty $ at least exponentially. Since $R(\pi) >0$, $R(\pi) -\pi -r\rightarrow \infty $, so that $\dot{\theta}\rightarrow
-\infty $, which is incompatible with $\theta >\theta _{\min }$. The
solution therefore cannot be extended to all $t\geq 0$: Output reaches $\bar{%
y}$ in finite time, where $\Gamma \rightarrow \infty $.

\textit{Monotonicity of }$V(\cdot) $\textit{.} Let $\pi ^{\prime }(0) >\pi (0) $ be two initial inflation rates on the stable arm. If
they lie on the same branch, the trajectory starting at $\pi ^{\prime }(0) $
reaches $\pi (0) $ at some date $\tau >0$, by monotonicity; on the upper
branch, inflation is decreasing over time, so that $\pi ^{\prime }(t) =\pi (t-\tau) >\pi (t) $ for $t\geq \tau $%
, while $\pi ^{\prime }(t) \geq \pi (0) \geq \pi (t)
$ for $t<\tau $, with strict inequality for $t>0$; the lower branch is
analogous. If they lie on different branches, $\pi ^{\prime }(t)
>\pi ^{\ast }>\pi (t) $ for all $t$. In either case, since $%
R(\pi) -\pi $ is strictly decreasing, the real interest rate is
strictly lower along the trajectory starting at $\pi ^{\prime }(0) $ at every
$t>0$, the discount factor is strictly higher at every horizon, and, $\bar{s}%
(\cdot) $ being positive, $V(\pi ^{\prime }(0))
>V(\pi (0)) $.

\section*{Appendix E. Active Monetary Policy}

Let $F(\pi) \equiv R(\pi) -\pi -r$, which is
negative on $(\pi _{L},\pi ^{\ast }) $, positive outside $\left[
\pi _{L},\pi ^{\ast }\right] $, and diverges to $+\infty $ as $\pi
\rightarrow \pm \infty $, so that $\dot{\theta}=-F(\pi) $.
Assume that the unintended steady state exists, $\Gamma _{0}<r(\pi
_{L}-\pi ^{\ast }) $. I show that every solution defined for all $%
t\geq 0$ is bounded and converges either to the intended steady state, from
which it starts, or to the liquidity trap, along its stable arm.

\textit{Step 1.} By the essentiality of leisure, $\theta >\theta _{\min }$
along every solution.

\textit{Step 2: marginal utility is bounded above.} By the existence of the
unintended steady state, there exist $\epsilon >0$ and $\bar{\theta}$ such
that $\Gamma (y(\theta)) \leq r(\pi _{L}-\pi
^{\ast }) -\epsilon $ for $\theta \geq \bar{\theta}$. Marginal utility
rises only while $\pi \in (\pi _{L},\pi ^{\ast }) $. Whenever $%
\theta \geq \bar{\theta}$ and $\pi \in \left[ \pi _{L},\pi ^{\ast }\right] $%
, one has $\dot{\pi}\geq \epsilon $, so that the band is crossed upward in a
time not exceeding $(\pi ^{\ast }-\pi _{L}) /\epsilon $, and it
cannot be re-entered from above as long as $\theta \geq \bar{\theta}$, since
$\dot{\pi}=-\Gamma >0$ at $\pi =\pi ^{\ast }$. Hence $\theta (t)
\leq \theta _{\max }\equiv \max \left\{ \theta (0) ,\bar{\theta}%
\right\} +\bar{F}(\pi ^{\ast }-\pi _{L}) /\epsilon $ for all $t$,
where $\bar{F}\equiv \max_{\left[ \pi _{L},\pi ^{\ast }\right] }(-F) $.

\textit{Step 3: inflation is bounded above.} By Step 2, $\Gamma \geq -g$ for
some constant $g$, so that $\dot{\pi}\leq r(\pi -\pi ^{\ast })
+g $: Inflation can grow at most exponentially. Fix $\hat{\pi}>\pi ^{\ast }$
with $F(\hat{\pi}) =K>0$. While $\pi >\hat{\pi}$, $\dot{\theta}%
\leq -K$; since $\theta \in \left( \theta _{\min },\theta _{\max }\right] $,
each episode above $\hat{\pi}$ lasts at most $T\equiv (\theta _{\max
}-\theta _{\min }) /K$, during which $\pi -\pi ^{\ast }+g/r$ grows at
most by the factor $e^{rT}$. Hence $\pi (t) \leq \max \left\{
\pi (0) ,\pi ^{\ast }-g/r+(\hat{\pi}-\pi ^{\ast
}+g/r) e^{rT}\right\} $. This is the step that excludes explosive
inflationary paths: The rising real interest rate drives marginal utility
toward its lower bound too fast for inflation to grow without limit.

\textit{Step 4: inflation is bounded below.} By Step 2, $\Gamma \geq \Gamma
(y(\theta _{\max })) $. If $\pi <\pi ^{\ast
}+\Gamma (y(\theta _{\max })) /r$, then $\dot{\pi}%
<0$ forever, $\pi \rightarrow -\infty $, $F\rightarrow \infty $, and $\theta
\rightarrow -\infty $, contradicting Step 1.

\textit{Step 5: marginal utility is bounded away from its lower bound.} If $%
\theta $ approached $\theta _{\min }$, $\Gamma $ would become so large that
inflation would fall below $\pi _{L}$ before marginal utility could rise
appreciably (within the band, $\dot{\theta}\leq \bar{F}$); below $\pi _{L}$,
$\theta $ decreases and $\Gamma $ increases, so that $\dot{\pi}<0$ forever,
contradicting Step 4.

\textit{Step 6.} The trajectory thus remains in a compact subset of the
domain. By the Poincar\'{e}--Bendixson theorem and the absence of closed
orbits and closed chains of trajectories (Appendix C), its limit set is a
steady state. The intended steady state, a source, can be the limit only of
the trajectory that starts on it; otherwise the limit is the liquidity trap,
approached along its stable arm.

As a by-product, every trajectory that is not defined for all $t\geq 0$
enters, before ceasing to exist, the region $D_{L}\equiv \left\{ \pi <\pi
_{L},\dot{\pi}<0\right\} $: By Steps 2-4, applied on the maximal interval of
existence, a solution can cease to exist only as $\theta \rightarrow \theta
_{\min }$, where $\Gamma \rightarrow \infty $ and inflation falls below $\pi
_{L}$ with $\dot{\pi}<0$. The region $D_{L}$ is positively invariant, by the
same argument used for $D$ in Appendix D, since $\dot{\theta}<0$ below $%
\pi _{L}$. Every trajectory of the real block therefore either converges to
a steady state or enters $D_{L}$.

\section*{Appendix F. Heteroclinic Connection}

Take a small circle $C$ around the intended steady state, small enough that,
backward in time, every trajectory starting on $C$ converges to the intended
steady state, which is possible since the latter is a source. No point of $C$
converges forward to the intended steady state, so that, by Appendix E,
each point of $C$ either lies on the stable arm of the liquidity trap or
enters $D_{L}$. Suppose that every point of $C$ entered $D_{L}$. Since
entering an open set is a property that persists under small perturbations
of the initial condition, and $C$ is compact, each point of $C$ can be
followed along its trajectory, continuously in the initial point, until it
lies in $D_{L}$.\footnote{%
Formally, for each $x\in C$ there exist $t_{x}$ and a neighborhood $U_{x}$
such that the flow maps $U_{x}$ into $D_{L}$ at time $t_{x}$. By compactness,
finitely many such neighborhoods cover $C$, and a partition of unity
subordinate to the cover delivers a continuous stopping time; since $D_{L}$
is positively invariant, every trajectory is in $D_{L}$ at all intermediate
times.} The flow thereby deforms $C$ continuously into a closed curve lying
entirely below $\pi _{L}$. Since no trajectory can reach the intended steady
state, a fixed point, the deformed curve must still encircle it.\footnote{%
This is the invariance of the winding number of a closed curve around a
point under continuous deformations that avoid the point (e.g., Hirsch,
Smale and Devaney, 2013).} This is impossible for a curve lying entirely
below $\pi _{L}<\pi ^{\ast }$. Hence some point of $C$ lies on the stable arm
of the liquidity trap. Backward in time this point converges to the intended
steady state, by the choice of $C$, so that its trajectory is a heteroclinic
connection from the intended steady state to the liquidity trap. As $C$ can
be chosen arbitrarily small, the continuum of equilibria converging to the
liquidity trap includes trajectories originating arbitrarily close to the
intended steady state. The argument uses only the boundedness established in
Appendix E and the absence of closed orbits established in Appendix C:
Unlike the bifurcation argument of Benhabib, Schmitt-Groh\'{e} and Uribe
(2001a, Proposition 1), it is not confined to parameter values close to a
bifurcation point, and it leaves no room for limit cycles.

\section*{Appendix G. Numerical Illustration}

The figures are drawn from a numerical illustration with quarterly
parameters $r=0.01$ and $\pi ^{\ast }=0.005$ (a 4 percent real interest rate
and a 2 percent inflation target in annual terms), $u(c)
=c^{1-\sigma }/(1-\sigma) $ with $\sigma =2$, $f(h)
=h^{0.7}$, $z(h) =-h-\ln (1-h) $ with $\bar{h}=1$%
---the logarithmic utility of leisure augmented with a linear term that
sets the marginal disutility of labor to zero at $h=0$---and $\eta =-6$
(a 20 percent markup). The degree of price stickiness $%
\gamma $ is set so that the oscillations of the real block around the
intended steady state are visible in Figure 1; the qualitative features of
the figures do not depend on it. The active rule is the exponential rule described in subsection 2.2
with $A=3$, which delivers an unintended steady state with
annual deflation of about 3.6 percent and an annual nominal interest rate of
about 0.4 percent. The passive rule is $R(\pi) =(\varphi /k) \ln \left\{ 1+\exp \left[
k(\pi -\xi) \right] \right\} $, which is convex, approaches zero as $\pi
\rightarrow -\infty $, and is asymptotic to a straight line with slope
$\varphi $ as $\pi \rightarrow \infty $; I set $\varphi =0.8$ and choose $k$
and $\xi $ so that $R(\pi ^{\ast }) =r+\pi ^{\ast }$ and $R^{\prime }(\pi
^{\ast }) =0.5$. Figure 3 is drawn for $\alpha =0$ and a constant
exogenous surplus.

\section*{References}

\begin{description}
\item[\hspace{-1cm}] Andr\'{e}s, J., J. D. L\'{o}pez-Salido and J. Vall\'{e}s
(2006), \textquotedblleft Money in an Estimated Business Cycle Model of the
Euro Area,\textquotedblright\ \textit{Economic Journal} 116, 457-477.%
\vspace{-0.1in}

\item[\hspace{-1cm}] Ascari, G. and L. Rossi (2012), \textquotedblleft Trend
Inflation and Firms Price-Setting: Rotemberg Versus Calvo,\textquotedblright\
\textit{Economic Journal} 122, 1115-1141.\vspace{-0.1in}

\item[\hspace{-1cm}] Ascari, G. and A. M. Sbordone (2014), \textquotedblleft
The Macroeconomics of Trend Inflation,\textquotedblright\ \textit{Journal of
Economic Literature} 52, 679-739.\vspace{-0.1in}

\item[\hspace{-1cm}] Bassetto, M. (2002), \textquotedblleft A Game-Theoretic
View of the Fiscal Theory of the Price Level,\textquotedblright\ \textit{%
Econometrica} 70, 2167-2195.\vspace{-0.1in}

\item[\hspace{-1cm}] Bassetto, M. and T. J. Sargent (2020),
\textquotedblleft Shotgun Wedding: Fiscal and Monetary
Policy,\textquotedblright\ \textit{Annual Review of Economics} 12, 659-690.%
\vspace{-0.1in}

\item[\hspace{-1cm}] {Benhabib, J., S. Schmitt-Groh\'{e}\ and M. Uribe
(2001a)}, \textquotedblleft The Perils of Taylor Rules,\textquotedblright\
\textit{Journal of Economic Theory} 96, 40-69.\vspace{-0.1in}

\item[\hspace{-1cm}] {Benhabib, J., S. Schmitt-Groh\'{e}\ and M. Uribe
(2001b), \textquotedblleft Monetary Policy and Multiple
Equilibria,\textquotedblright\ \textit{American Economic Review }}91,
167-186.\vspace{-0.1in}

\item[\hspace{-1cm}] {Benhabib, J., S. Schmitt-Groh\'{e}\ and M. Uribe
(2002), }\textquotedblleft Avoiding Liquidity Traps,\textquotedblright\
\textit{Journal of Political Economy} 110, 535-563.\vspace{-0.1in}

\item[\hspace{-1cm}] Bianchi, F. (2012), \textquotedblleft Evolving
Monetary/Fiscal Policy Mix in the United States,\textquotedblright\
\textit{American Economic Review} 102(3), 167-172.\vspace{-0.1in}

\item[\hspace{-1cm}] Bianchi, F. (2013), \textquotedblleft Regime Switches,
Agents' Beliefs, and Post-World War II U.S. Macroeconomic
Dynamics,\textquotedblright\ \textit{Review of Economic Studies} 80,
463-490.\vspace{-0.1in}

\item[\hspace{-1cm}] Blanchard, O. J. and C. M. Kahn (1980),
\textquotedblleft The Solution of Linear Difference Models under Rational
Expectations,\textquotedblright\ \textit{Econometrica} 48, 1305-1311.%
\vspace{-0.1in}

\item[\hspace{-1cm}] Bohn, H. (1995), \textquotedblleft The Sustainability
of Budget Deficits in a Stochastic Economy,\textquotedblright\ \textit{%
Journal of Money, Credit and Banking} 27, 257-271.\vspace{-0.1in}

\item[\hspace{-1cm}] Bohn, H. (1998), \textquotedblleft The Behavior of U.S.
Public Debt and Deficits,\textquotedblright\ \textit{Quarterly Journal of
Economics} 113, 949-963.\vspace{-0.1in}

\item[\hspace{-1cm}] Brock, W. A. (1974), \textquotedblleft Money and
Growth: The Case of Long Run Perfect Foresight,\textquotedblright\
\textit{International Economic Review} 15, 750-777.\vspace{-0.1in}

\item[\hspace{-1cm}] Brock, W. A. (1975), \textquotedblleft A Simple Perfect
Foresight Monetary Model,\textquotedblright\ \textit{Journal of Monetary
Economics} 1, 133-150.\vspace{-0.1in}

\item[\hspace{-1cm}] Calvo, G. A. (1983), \textquotedblleft Staggered Prices
in a Utility-Maximizing Framework,\textquotedblright\ \textit{Journal of
Monetary Economics} 12, 383-398.\vspace{-0.1in}

\item[\hspace{-1cm}] Canzoneri, M., R. Cumby and B. Diba (2010),
\textquotedblleft The Interaction Between Monetary and Fiscal
Policy,\textquotedblright\ in B. M. Friedman and M. Woodford (eds), \textit{%
Handbook of Monetary Economics} 3, Amsterdam/Boston: North-Holland/Elsevier,
935-999.\vspace{-0.1in}

\item[\hspace{-1cm}] Cochrane, J. H. (1998), \textquotedblleft A
Frictionless View of U.S. Inflation,\textquotedblright\ \textit{NBER
Macroeconomics Annual}, 323-421.\vspace{-0.1in}

\item[\hspace{-1cm}] Cochrane, J. H. (2005), \textquotedblleft Money as
Stock,\textquotedblright\ \textit{Journal of Monetary Economics} 52, 501-528.%
\vspace{-0.1in}

\item[\hspace{-1cm}] Cochrane, J. H. (2011), \textquotedblleft Determinacy
and Identification with Taylor Rules,\textquotedblright\ \textit{Journal of
Political Economy} 119, 565-615.\vspace{-0.1in}

\item[\hspace{-1cm}] Cochrane, J. H. (2017), \textquotedblleft The
New-Keynesian Liquidity Trap,\textquotedblright\ \textit{Journal of Monetary
Economics} 92, 47-63.\vspace{-0.1in}

\item[\hspace{-1cm}] Cochrane, J. H. (2022a), \textquotedblleft The Fiscal
Roots of Inflation,\textquotedblright\ \textit{Review of Economic Dynamics}
45, 22-40.\vspace{-0.1in}

\item[\hspace{-1cm}] Cochrane, J. H. (2022b), \textquotedblleft Fiscal
Histories,\textquotedblright\ \textit{Journal of Economic Perspectives}
36(4), 125-146.\vspace{-0.1in}

\item[\hspace{-1cm}] Cochrane, J. H. (2023), \textit{The Fiscal Theory of
the Price Level}, Princeton: Princeton University Press.\vspace{-0.1in}

\item[\hspace{-1cm}] Eggertsson, G. B. and M. Woodford (2003),
\textquotedblleft The Zero Bound on Interest Rates and Optimal Monetary
Policy,\textquotedblright\ \textit{Brookings Papers on Economic Activity}
2003(1), 139-211.\vspace{-0.1in}

\item[\hspace{-1cm}] {Gal\'{\i}, J. (2015), \textit{Monetary Policy,
Inflation, and the Business Cycle}, 2nd ed., Princeton: Princeton University
Press.}\vspace{-0.1in}

\item[\hspace{-1cm}] Hirsch, M. W., S. Smale and R. L. Devaney (2013),
\textit{Differential Equations, Dynamical Systems, and an Introduction to
Chaos}, 3rd ed., Amsterdam: Academic Press/Elsevier.\vspace{-0.1in}

\item[\hspace{-1cm}] Ireland, P. N. (2004), \textquotedblleft Money's Role in
the Monetary Business Cycle,\textquotedblright\ \textit{Journal of Money,
Credit and Banking} 36, 969-983.\vspace{-0.1in}

\item[\hspace{-1cm}] Kaplan, G. (2026), \textquotedblleft A Note on
Equilibrium Uniqueness in New Keynesian Models and Implications for the
Fiscal Theory of the Price Level,\textquotedblright\ \textit{Journal of
Macroeconomics} 90, 103814.\vspace{-0.1in}

\item[\hspace{-1cm}] Leeper, E. M. (1991), \textquotedblleft Equilibria
under `Active' and `Passive' Monetary and Fiscal
Policies,\textquotedblright\ \textit{Journal of Monetary Economics} 27,
129-147.\vspace{-0.1in}

\item[\hspace{-1cm}] Leeper, E. M. and C. Leith (2016), \textquotedblleft
Understanding Inflation as a Joint Monetary-Fiscal
Phenomenon,\textquotedblright\ in J. B. Taylor and H. Uhlig (eds), \textit{%
Handbook of Macroeconomics} 2, Amsterdam: North-Holland/Elsevier, 2305-2415.%
\vspace{-0.1in}

\item[\hspace{-1cm}] Leeper, E. M. and B. Li (2017), \textquotedblleft
Surplus-Debt Regressions,\textquotedblright\ \textit{Economics Letters} 151,
10-15.\vspace{-0.1in}

\item[\hspace{-1cm}] McCallum, B. T. (2001), \textquotedblleft Monetary
Policy Analysis in Models without Money,\textquotedblright\ \textit{Federal
Reserve Bank of St. Louis Review} 83, 145-164.\vspace{-0.1in}

\item[\hspace{-1cm}] Perko, L. (2001), \textit{Differential Equations and
Dynamical Systems}, 3rd ed., New York: Springer.\vspace{-0.1in}

\item[\hspace{-1cm}] Roberts, J. M. (1995), \textquotedblleft New Keynesian
Economics and the Phillips Curve,\textquotedblright\ \textit{Journal of
Money, Credit and Banking} 27, 975-984.\vspace{-0.1in}

\item[\hspace{-1cm}] Rotemberg, J. J. (1982), \textquotedblleft Sticky
Prices in the United States,\textquotedblright\ \textit{Journal of Political
Economy} 90, 1187-1211.\vspace{-0.1in}

\item[\hspace{-1cm}] Sargent, T. J. and N. Wallace (1981), \textquotedblleft
Some Unpleasant Monetarist Arithmetic,\textquotedblright\ \textit{Federal
Reserve Bank of Minneapolis Quarterly Review} 5(3), 1-17.\vspace{-0.1in}

\item[\hspace{-1cm}] Sidrauski, M. (1967), \textquotedblleft Rational
Choice and Patterns of Growth in a Monetary Economy,\textquotedblright\
\textit{American Economic Review} 57(2), 534-544.\vspace{-0.1in}

\item[\hspace{-1cm}] Sims, C. A. (1994), \textquotedblleft A Simple Model
for the Study of the Determination of the Price Level and the Interaction of
Monetary and Fiscal Policy,\textquotedblright\ \textit{Economic Theory} 4,
381-399.\vspace{-0.1in}

\item[\hspace{-1cm}] {Taylor}, J. B. (1993), \textquotedblleft Discretion
Versus Policy Rules in Practice,\textquotedblright\ \textit{%
Carnegie-Rochester Conference Series on Public Policy} 39, 195-214.\vspace{%
-0.1in}

\item[\hspace{-1cm}] {Taylor}, J. B. (Ed.) (1999), \textit{Monetary Policy
Rules}, Chicago: University of Chicago Press.\vspace{-0.1in}

\item[\hspace{-1cm}] Walsh, C. E. (2017), \textit{Monetary Theory and Policy}%
, 4th ed., Cambridge: The MIT Press.\vspace{-0.1in}

\item[\hspace{-1cm}] Werning, I. (2011), \textquotedblleft Managing a
Liquidity Trap: Monetary and Fiscal Policy,\textquotedblright\ NBER Working
Paper 17344.\vspace{-0.1in}

\item[\hspace{-1cm}] Woodford, M. (1994), \textquotedblleft Monetary Policy
and Price Level Determinacy in a Cash-in-Advance Economy,\textquotedblright\
\textit{Economic Theory} 4, 345-380.\vspace{-0.1in}

\item[\hspace{-1cm}] Woodford, M. (1995), \textquotedblleft Price Level
Determinacy without Control of a Monetary Aggregate,\textquotedblright\
\textit{Carnegie-Rochester Conference Series on Public Policy} 43, 1-46.%
\vspace{-0.1in}

\item[\hspace{-1cm}] Woodford, M. (2001), \textquotedblleft Fiscal
Requirements for Price Stability,\textquotedblright\ \textit{Journal of
Money, Credit and Banking} 33, 669-728.\vspace{-0.1in}

\item[\hspace{-1cm}] Woodford, M. (2003), \textit{Interest and Prices},
Princeton: Princeton University Press.\vspace{-0.1in}

\item[\hspace{-1cm}] Woodford, M. (2008), \textquotedblleft How Important Is
Money in the Conduct of Monetary Policy?\textquotedblright\ \textit{Journal
of Money, Credit and Banking} 40, 1561-1598.\vspace{-0.1in}
\end{description}

\end{doublespace}

\end{document}